\documentstyle[aps,pre,eqsecnum,psfig,epsfig]{revtex}

\def\eps{\varepsilon}
\def\Dm{\widetilde{\cal D}_{\mu}}
\def\E{\overline{\cal E}}
\def\D{{\cal D}}
\def\k{{\bf k}}
\def\n{{\bf n}}
\def\p{{\bf p}}
\def\q{{\bf q}}
\def\bfkap{\mbox{\boldmath $\kappa$}}

\begin{document}
\draft


\title{Renormalization-group approach to the stochastic Navier--Stokes
equation:  \\ Two-loop approximation}

\author{L.Ts.~Adzhemyan, N.V.~Antonov, M.V.~Kompaniets, and A.N.~Vasil'ev}
\address{Department of Theoretical Physics, St.~Petersburg University,
Uljanovskaja 1, St.~Petersburg, Petrodvorez, 198504 Russia}


\maketitle

\begin{abstract}
The field theoretic renormalization group is applied to the stochastic
Navier--Stokes equation that describes fully developed fluid turbulence.
The complete two-loop calculation of the renormalization constant, the
$\beta$ function, the fixed point and the ultraviolet correction exponent
is performed. The Kolmogorov constant and the inertial-range skewness
factor, derived to second order of the $\eps$ expansion, are in a good
agreement with the experiment. The possibility of the extrapolation of
the $\eps$ expansion beyond the threshold where the sweeping effects
become important is demonstrated on the example of a Galilean-invariant
quantity, the equal-time pair correlation function of the velocity field.
The extension to the $d$-dimensional case is briefly discussed.
\end{abstract}


\section{Introduction} \label{sec:Intro}

One of the oldest open problems in theoretical physics is that of describing
fully developed turbulence on the basis of a microscopic model. The latter
is usually taken to be the stochastic Navier--Stokes (NS) equation subject
to an external random force that models the energy injection by the
large-scale modes; see, e.g., \cite{Monin,Legacy}. The aim of the theory is
to verify the basic principles of the celebrated Kolmogorov--Obukhov
phenomenological theory, study deviations from this theory, determine
the dependence of various correlation functions on the times, distances,
external (integral) and internal (viscous) turbulence scales, and derive
the corresponding scaling dimensions. Most results of this kind were obtained
within the framework of numerous semiphenomenological models that cannot be
considered to be the basis for construction of a regular expansion in certain
small (at least formal) parameter \cite{Monin,Legacy}.

An important exception is provided by the renormalization group (RG) method
that was earlier successfully applied in the theory of critical behavior
to explain the origin of critical scaling and to calculate universal
quantities (critical dimensions and scaling functions) in the form of the
$\eps$ expansions \cite{Zinn,book}.

The RG approach to the stochastic NS equation, pioneered in
\cite{Nelson,Dominicis,Frisch,Pismak}, allows one to prove the existence
of the infrared (IR) scale invariance with exactly known ``Kolmogorov''
dimensions and the independence of the correlation functions of the viscous
scale (the second Kolmogorov hypothesis), and calculate a number of
representative constants in a reasonable agreement with experiment.
Detailed review of the RG theory of turbulence and more references can
be found in \cite{UFN,turbo}.

The standard RG formalism is applicable to the stochastic NS equation if
the correlation function of the random force is chosen in the form
$\propto k^{4-d-2\eps}$, where $k$ is the momentum (wave number),
$d$ is the
space dimension and the exponent $\eps$ plays the part analogous to that
played by $4-d$ in the RG theory of critical behavior. Although the results
of the RG analysis are reliable and internally consistent for small $\eps$,
the possibility of their extrapolation to the real value $\eps=2$ (see
below) and thus their relevance for the real fluid turbulence was called in
question in a number of studies, e.g. \cite{K,CK,Teo,Teo2,Woo}.

The crucial role in recent studies of intermittency and anomalous scaling of
fluid turbulence was played by a simple model of a passive scalar quantity
advected by a random Gaussian field, white in time and self-similar in space,
the so-called Kraichnan's rapid-change model \cite{Kraich1}.
There, for the first time the existence of anomalous
scaling was established on the basis of a microscopic model \cite{Kraich2}
and the corresponding anomalous exponents were calculated within controlled
approximations \cite{GK,Falk1} and numerical experiments \cite{VMF}.
Detailed review of the recent theoretical research on the passive scalar
problem and more references can be found in \cite{FGV}.

The field theoretic RG and the operator product
expansion (OPE) were applied to the rapid-change model in \cite{RG}.
The RG allows one to construct a systematic perturbation expansion for
the anomalous exponents, analogous to the $\varepsilon$ expansion in models
of critical behavior, and to calculate the exponents to the second
\cite{RG} and third \cite{cube} orders.
For passively advected vector fields
the exponents for higher-order correlation functions were derived by
means of the RG techniques to the leading order in $\eps$
in Refs. \cite{vector}.
It was shown that the knowledge of three
terms allows one to obtain reasonable predictions for finite $\eps\sim1$;
even the plain $\eps$ expansion captures some subtle qualitative features
of the anomalous exponents established in numerical experiments \cite{cube}.

In contrast to standard $\phi^{4}$ model of critical behavior
\cite{Zinn,book}, where the critical exponents are known up to the order
$\eps^{5}$ (five-loop approximation), and to the rapid-change model,
where the anomalous exponents are known up to $\eps^{3}$,
all the calculations in the RG
approach to the stochastic NS equation have been confined with the simplest
one-loop approximation. The reason for this distinction is twofold.
First, any multiloop calculation for this dynamical model is a demanding
job: one can say that the two-loop calculation for the stochastic NS
equation is as cumbersome as the four-loop calculation for the conventional
$\phi^{4}$ model. Second, the critical dimensions for the most important
physical quantities (velocity and its powers, frequency, energy dissipation
rate and so on) are given by the one-loop approximation exactly (the
corresponding $\eps$ series terminate at first-order terms) and the
higher-order calculations for them are not needed.

However, the $\eps$ series for other important quantities do not terminate
and the calculation of the higher-order terms for them is of great interest.
In this paper, we present the results of the two-loop calculation for a
number of such quantities: the $\beta$ function, the coordinate of the RG
fixed point, the ultraviolet (UV) correction exponent $\omega$, the
Kolmogorov constant $C_{K}$ and the inertial-range skewness factor
${\cal S}$. The knowledge of  higher-order terms is also important
to judge about the validity and convergence properties of the $\eps$
expansions in the models of  turbulence on the whole.

The plan of the paper is as follows. In Sec.~\ref{sec:Model} we introduce
the stochastic NS equation paying special attention to the choice of the
random force correlator and the physical value of the parameter $\eps$.
In Sec.~\ref{sec:QFT} we recall the field theoretic formulation of the
model, diagrammatic technique, renormalization and RG equations.
Since the ``ideology''  of the RG and OPE approach to the stochastic NS
model is explained in Refs. \cite{UFN,turbo,JETP} in detail, here we confine
ourselves to only the necessary information. In Sec.~\ref{sec:ZZ} we
present in detail the two-loop calculation of the renormalization
constant, RG functions, coordinate of the fixed point and the UV
correction exponent. We explain some tricks (e.g. an appropriate
choice of the IR cutoff) that essentially simplify the calculation
and make it feasible.

In Sec.~\ref{sec:pair} we calculate two orders of the $\eps$ expansion
for the pair correlation function of the velocity field (this accuracy is
consistent with the two-loop calculation of the RG functions in
Sec.~\ref{sec:ZZ}). They are used later in Sec.~\ref{sec:CK} in the
calculation of the skewness factor and the Kolmogorov constant.

In Sec.~\ref{sec:bsk} we discuss on the example of the pair correlation
function the possibility of the extrapolation of the $\eps$ expansion
to finite values of $\eps$. We explain an important distinction that
exists between the RG technique and the traditional approach based on
diagrammatic self-consistency equations and argue that, contrary to what
is sometimes claimed, the RG is applicable beyond the threshold where the
so-called sweeping effects become important, but one should combine the
RG with the OPE and go beyond the plain $\eps$ expansions.

In Sec.~\ref{sec:CK} we perform the two-loop calculation of the Kolmogorov
constant $C_{K}$ and the inertial-range skewness factor ${\cal S}$.
The new point is not only the inclusion of the second-order correction, but
also the derivation of $C_{K}$ through an universal (in the sense of the
theory of critical behavior) quantity. This allows us to circumvent the
main drawback of earlier calculations of $C_{K}$: the intrinsic ambiguities
in the corresponding $\eps$ expansions. To the best of our knowledge, the
{\it inertial-range} skewness factor ${\cal S}$ has never been calculated
earlier within the RG approach.

The experience on the RG theory of critical behavior shows that for
dynamical models the higher-order corrections are not small, and in order
to obtain reasonable predictions, say, for critical exponents, one should
augment the plain $\eps$ expansions by additional information obtained
from the instanton calculus and Pad\'e--Borel summations; the situation for
the {\it amplitudes} in scaling laws is even worse (see e.g. \cite{Zinn}).
Surprisingly enough, in our case already the first-order results are in a
reasonable agreement with experiment, which is not destroyed by the inclusion
of the second-order corrections. It turns out that the experimental value
of $C_{K}$ and ${\cal S}$ lie in between of the two consecutive
approximations, like for the exactly solvable Heisenberg model \cite{Hei}.

The main conclusion of the paper is that the renormalization group and the
$\eps$ expansion allow one to derive a number of important characteristics
of the real fluid turbulence beyond the simplest first-order approximations
and in a good agreement with the experiment.

\section{Stochastic Navier--Stokes equation and the choice
of the random force} \label{sec:Model}

As the microscopic model of the fully developed, homogeneous, isotropic
turbulence of an incompressible viscous fluid one usually takes the
stochastic NS equation with a random driving force
\begin{equation}
\nabla _t\varphi _i=\nu _0\partial^{2} \varphi _i-\partial _i
{\cal P}+f_i ,
\qquad
\nabla _t\equiv \partial _t+(\varphi \partial)  .
\label{1.1}
\end{equation}
Here $\varphi_i$ is the transverse (due to the incompressibility) vector
velocity field, ${\cal P}$ and $f_i$ are the pressure and the transverse
random force per unit mass (all these quantities depend on
$x\equiv\{t,{\bf x}\}$), $\nu _0$ is the kinematical viscosity coefficient,
$\partial^{2}$ is the Laplace operator and $\nabla_t$ is the Lagrangian
derivative. The problem (\ref{1.1}) is studied on the entire $t$ axis and
is augmented by the retardation condition and the condition that
$\varphi_i$ vanishes for $t\to-\infty$. We assume for $f$ a Gaussian
distribution with zero average and correlator
\begin{equation}
\big\langle f_i(x)f_j(x')\big\rangle = \frac{\delta (t-t')}{(2\pi)^{d}}\,
\int d{\bf k}\, P_{ij}({\bf k})\, d_f(k)\, \exp \big[{\rm i}{\bf k}
\left({\bf x}-{\bf x}'\right)\big] ,
\label{1.2}
\end{equation}
where $P_{ij}({\bf k}) =\delta _{ij}  - k_i k_j / k^2$ is the transverse
projector, $d_f(k)$ is some function of $k\equiv |{\bf k}|$ and model
parameters, and $d$ is the dimension of the ${\bf x}$ space. The time
decorrelation of the random force guarantees Galilean invariance of the
full stochastic problem (\ref{1.1}), (\ref{1.2}).

Let us specify the form of the function $d_f(k)$ in the correlator
(\ref{1.2}) used in the RG theory of turbulence; more detailed
discussion is given in Refs. \cite{UFN,turbo}. Physically, the random force
models the injection of energy to the system owing to interaction with
the large-scale eddies. Idealized injection by infinitely large eddies
corresponds to $d_{f}(k) \propto \delta({\bf k})$, more precisely,
\begin{equation}
d_f(k)= 2 (2\pi)^{d}\, \E\, \delta({\bf k}) / (d-1),
\label{2.75}
\end{equation}
where $\E$ is the average power of the injection (equal to the average
dissipation rate) and the amplitude factor comes from the exact relation
\begin{equation}
\E = \frac{(d-1)}{2 (2\pi)^{d}} \,\int d {\bf k} \, d_f(k).
\label{1.3}
\end{equation}
On the other hand, for the use of the standard RG technique it is important
that the function $d_{f}(k)$ have a power-law behavior at large $k$. This
condition is satisfied if $d_{f}(k)$ is chosen in the form \cite{Dominicis}
\begin{equation}
d_f(k)=D_0\,k^{4-d-2\varepsilon},
\label{1.10}
\end{equation}
where $D_0>0$ is the amplitude factor and $\varepsilon>0$ is the exponent
with the physical value $\varepsilon=2$ (see below).

Let us recall the well-known power-law representation of the $d$-dimensional
$\delta$ function,
\begin{eqnarray}
\delta({\bf k}) = \lim_{\eps\to 2}\, \frac{1}{(2\pi)^{d}} \int d{\bf x}
\, (\Lambda x)^{2\eps-4} \, \exp [{\rm i} ({\bf k}\cdot{\bf x})] = S_{d}^{-1}
k^{-d} \lim_{\eps\to 2} \left[ (4-2\eps) (k/\Lambda)^{4-2\eps} \right],
\label{2.76}
\end{eqnarray}
with some ultraviolet (UV) momentum scale $\Lambda$. Here and below we denote
\begin{eqnarray}
S_d \equiv 2\pi^{d/2}/\Gamma (d/2), \qquad \bar S_d \equiv S_d / (2\pi)^{d},
\label{surface}
\end{eqnarray}
where $S_d$ is the surface area of the unit sphere in $d$-dimensional space
and $\Gamma(\cdots)$ is Euler's Gamma function.
It then follows that for $\eps\to2$, the function (\ref{1.10}) turns to the
ideal injection (\ref{2.75}) if the amplitude $D_{0}$ is related to $\E$ as
\begin{eqnarray}
D_{0} \to \frac{4(2-\eps)\,\Lambda^{2\eps-4}}
{\overline S_{d}(d-1)}\,\, \E \qquad {\rm for} \ \eps\to2.
\label{2.74}
\end{eqnarray}

A more realistic model, used e.g. in \cite{Dominicis,JETP}, is
\begin{equation}
d_f(k)=D_0\,k^{4-d-2\varepsilon}\,h(m/k), \qquad  h(0)=1,
\label{1.9}
\end{equation}
where $m=1/L$ is the reciprocal of the integral turbulence scale $L$ and
$h(m/k)$ is some well-behaved function that provides the IR regularization.
Its specific form can be chosen to simplify the practical calculation in
higher orders; see Sec.~\ref{sec:ZZ}.

In the RG approach to the problem (\ref{1.1}), (\ref{1.2}), (\ref{1.9}) the
exponent $\varepsilon$ plays the part analogous to that played by $4-d$ in
Wilson's theory of critical phenomena \cite{Zinn,book}. All the results of
the RG analysis are reliable and internally consistent for small $\eps$; the
possibility of their extrapolation to the real value $\eps=2$ was called in
question in a number of studies, e.g. \cite{K,CK,Teo,Teo2,Woo}.
We shall return
to this important issue later, and here we only note that the replacement
of the ideal injection (\ref{2.75}) by the power-law model (\ref{1.10})
followed by the use of the $\eps$ expansion was confirmed by the example
of the exactly solvable Heisenberg model \cite{Hei}.

\section{Field theoretic formulation and renormalization of the model}
\label{sec:QFT}

Detailed exposition of the RG theory of turbulence and the bibliography
can be found in \cite{UFN,turbo}; below we restrict ourselves to only the
necessary information.

Stochastic problem (\ref{1.1}), (\ref{1.2}) is equivalent to the field
theoretic model of the doubled set of transverse vector fields
$\Phi\equiv\{\varphi,\varphi'\}$ with action functional
\begin{equation}
S(\Phi )=\varphi 'D_f\varphi '/2+\varphi '[-\partial _t\varphi +\nu
_0\partial^{2} \varphi -(\varphi \partial )\varphi ] ,
\label{action}
\end{equation}
where $D_f$ is the random force correlator (\ref{1.2}) and the required
integrations over $x=\{t,{\bf x}\}$ and summations over the vector indices
are understood.

The model (\ref{action}) corresponds to a standard diagrammatic perturbation
theory with bare propagators
\begin{eqnarray}
\langle \varphi \varphi '\rangle _0=\langle \varphi '\varphi \rangle _0^*
=(-{\rm i}\omega +\nu _0k^2)^{-1}, \quad
\langle \varphi '\varphi '\rangle _0=0, \quad
\langle \varphi \varphi \rangle _0=d_f(k)/(\omega ^2+\nu _0^2k^4)
\label{lines}
\end{eqnarray}
in the frequency-momentum ($\omega$--${\bf k}$) representation or
\begin{eqnarray}
\langle \varphi(t) \varphi '(t')\rangle _0= \theta(t-t')
\exp \left[-\nu_0 k^{2} (t-t') \right], \quad
\langle \varphi'(t)\varphi'(t')\rangle _0=0, \quad
\langle \varphi \varphi \rangle _0=d_f(k)
\exp \left[-\nu_0 k^{2} |t-t'| \right]\, / 2\nu_0 k^{2}
\label{lines2}
\end{eqnarray}
in the time-momentum ($t$--${\bf k}$) representation; the common
factor $P_{ij}({\bf k})$ in (\ref{lines}), (\ref{lines2}) is understood.
The interaction in (\ref{action}) corresponds to the triple vertex
$-\varphi'(\varphi\partial)\varphi=\varphi'_iV_{ijs}\varphi_j\varphi _s/2$
with vertex factor $V_{ijs}={\rm i}(k_j\delta _{is}+k_s\delta _{ij})$,
where ${\bf k}$ is the momentum of the field $\varphi'$. The part of the
coupling constant (expansion parameter in the ordinary perturbation theory)
is played by $g_0\equiv D_0/\nu _0^3$ with $D_0$ from (\ref{1.9}).

The model (\ref{action}) is logarithmic (the coupling constant $g_0$ is
dimensionless) at $\eps=0$, and the UV divergences have the form of the
poles in $\eps$ in the correlation functions of the fields
$\Phi\equiv\{\varphi,\varphi'\}$. Superficial UV divergences, whose removal
requires counterterms, are present only in the 1-irreducible function
$\langle\varphi'\varphi\rangle$, and the corresponding counterterm reduces
to the form $\varphi'\partial^{2}\varphi$. In the special case $d=2$ a new
UV divergence appears in the 1-irreducible function
$\langle\varphi'\varphi'\rangle$. We shall study this important case in a
separate paper, and for now we always assume $d>2$. Then for the
complete elimination of the UV divergences it is sufficient to perform
the multiplicative renormalization of the parameters $\nu_0$ and $g_{0}$
with the only independent renormalization constant $Z_{\nu}$:
\begin{equation}
\nu_0=\nu Z_{\nu}, \qquad g_{0}=g\mu^{2\eps}Z_{g},
\qquad Z_{g}=Z_{\nu}^{-3} \qquad (D_{0} = g_{0}\nu_0^{3}
= g\mu^{2\eps} \nu^{3}).
\label{18}
\end{equation}
Here $\mu$ is the reference mass in the minimal subtraction (MS) scheme,
which we always use in what follows, $g$ and $\nu$ are renormalized analogs
of the bare parameters $g_{0}$ and $\nu_0$, and $Z=Z(g,\eps,d)$ are the
renormalization constants.
No renormalization of the fields and the ``mass'' $m_{0}=m$ is needed,
i.e., $Z_{\Phi}=1$ for all $\Phi$ and $Z_{m}=1$.

In the MS scheme the renormalization
constants have the form ``1 + only poles in $\eps$,'' in particular,
\begin{eqnarray}
Z_{\nu}=1+\sum _{k=1}^{\infty }a_k(g)\eps ^{-k}=1+\sum _{n=1}^{\infty }g^n
\sum _{k=1}^{n}a_{nk}\eps ^{-k},
\label{1.30}
\end{eqnarray}
where the coefficients $a_{nk}$ depend only on $d$. The one-loop result
\begin{eqnarray}
a_{11}=-(d-1)\bar S_{d}/8(d+2),
\label{a11}
\end{eqnarray}
with $\bar S_{d}$ from (\ref{surface}), was presented in \cite{Pismak}.

Since the fields are not renormalized, their renormalized correlation
functions $W^{R}$ coincide with their unrenormalized analogs
$W=\langle\Phi\dots\Phi\rangle$; the only difference is in the choice of
variables and in the form of perturbation theory (in $g$ instead of $g_{0}$):
$W^{R} (g,\nu,\mu,m,\dots) = W (g_{0},\nu_0,m_{0},\dots)$. Here the dots
stand for other arguments like coordinates, times, momenta and so on. We
use $\Dm$ to denote the differential operator $\mu\partial_{\mu}$ for fixed
bare parameters $g_{0},\nu_0,m_{0}$ and operate on both sides of that
relation with it. This gives the basic differential RG equation:
\begin{equation}
{\cal D}_{RG}W^{R} (g,\nu,\mu,m,\dots) = 0, \quad
{\cal D}_{RG}\equiv {\cal D}_{\mu} + \beta(g)\partial_{g}
-\gamma_{\nu}(g){\cal D}_{\nu},
\label{RGE}
\end{equation}
where ${\cal D}_{RG}$ is the operation $\Dm$ expressed in
renormalized variables,
${\cal D}_{x}\equiv x\partial_{x}$ for any variable
$x$, and the RG functions (the anomalous dimension $\gamma_{\nu}$ and
the $\beta$ function) are defined as
\begin{equation}
\gamma_\nu(g) \equiv \Dm \ln Z_{\nu} = -2 {\cal D}_{g}  a_1(g),
\qquad
\beta(g,\varepsilon) \equiv \Dm g =
g\left[-2\varepsilon+3\gamma_\nu(g)\right]
\label{RGF1}
\end{equation}
with $a_1(g)$ from (\ref{1.30}). In the MS scheme only the residues at
the first-order poles in $\eps$, that is, only the coefficients $a_{k1}$,
contribute to the RG functions owing to the UV finiteness of the latter;
this explains the last relation for $\gamma_{\nu}$. The relation between
$\beta$ and $\gamma_{\nu}$ results from the definitions and the last
relation in (\ref{18}).

From the relations (\ref{RGF1}) using Eq. (\ref{a11}) we find the
first-order expressions for the RG functions:
\begin{equation}
\gamma_{\nu}(g)= (d-1) \bar S_{d}\, g  /4(d+2) +O(g^2),
\quad
\beta(g,\eps)= g\left[-2\eps+3(d-1)\bar S_{d}\,g/4(d+2)\right]+O(g^3).
\label{26}
\end{equation}
From (\ref{26}) it follows that an IR-attractive fixed point
\begin{equation}
g_* = 8(d+2)\eps\, /\, 3(d-1)\bar S_{d} +O(\eps^{2}),
\quad
\beta (g_*) = 0,
\quad
\omega\equiv\beta'(g_*)=2\eps +O(\eps^{2})>0
\label{FP1}
\end{equation}
of the RG equation (\ref{RGE}) exists in the physical region $g>0$ for
$\eps>0$. The value of $\gamma_{\nu}(g)$ at the fixed point is found
exactly using relations (\ref{RGF1}):
\begin{equation}
\gamma_{\nu}^{*} \equiv \gamma_{\nu}(g_*)= 2\eps/3 ,
\label{27}
\end{equation}
without corrections of order $\eps^{2}$, $\eps^{3}$, and so on.

For definiteness, consider the solution of the RG equation (\ref{RGE}) on the
example of a correlation function $W^{R}$ that involves $n$ fields $\varphi$
and depends on a single coordinate difference $r=|{\bf r}|$; the extension
to the general case is straightforward. We shall omit the superscript $R$ in
what follows. In renormalized variables, dimensionality considerations give:
\begin{equation}
W =  (\nu/r)^{n} R(s,u,g), \qquad s\equiv \mu r, \quad u\equiv mr,
\label{dimm}
\end{equation}
where $R(\cdots)$ is a function of completely dimensionless arguments (the
dependence on $d$ and $\eps$ is understood).

From the RG equation the identical representation follows,
\begin{equation}
W = (\nu/r)^{n} R(s,u,g) = (\bar\nu/r)^{n} R(1,u,\bar g),
\label{differ3}
\end{equation}
where the invariant variables $\bar e = \bar e(s,e)$ satisfy the
differential equation $\D_{RG} \bar e =0$ with the operator $\D_{RG}$ from
(\ref{RGE}) and the normalization conditions $\bar e = e$ at $s=1$
(we have used $e\equiv \{\nu,g,m\}$ to denote the full set of renormalized
parameters). The identity $\bar u \equiv u$ is a consequence of the
absence of ${\cal D}_m$ in the operator $\D_{RG}$ owing to the fact that
$m$ is not renormalized. Equation (\ref{differ3}) is valid because
both sides of it satisfy the RG equation and coincide for $s=1$
owing to the normalization of the invariant variables.

The relation between the bare and invariant variables has the form
\begin{equation}
\nu_0=\bar \nu Z_{\nu}(\bar g),
\qquad
g_{0}=\bar g r^{-2\eps}Z_{g}(\bar g)
\label{exo1}
\end{equation}
with $Z_{\nu,g}$ from (\ref{18}). Equation (\ref{exo1}) determines
implicitly the invariant variables as functions of the bare parameters;
it is valid because both sides of it satisfy the RG equation, and because
Eq. (\ref{exo1}) at $s\equiv\mu r=1$ coincides with (\ref{18}) owing to
the normalization conditions.

It is well known that the behavior of the invariant charge $\bar g$ at
large $\mu r$ is governed by the IR stable fixed point: $\bar g\to g_{*}$
with $g_{*}$ from (\ref{FP1}). Then the large-$\mu r$ behavior of the
invariant viscosity $\bar\nu$ is found explicitly from Eq. (\ref{exo1})
and the last relation in (\ref{18}):
\[\bar\nu = \left[ D_{0} r^{2\eps} /\bar g \right]^{1/3}
\to \left[ D_{0} r^{2\eps} / g_{*} \right]^{1/3} \, . \]
Then for $s\to\infty$  and any fixed $u\equiv mr$ we obtain
\begin{equation}
W = (D_{0}/g_{*})^{n/3}\, r^{-n\Delta_{\varphi}} f(u),
\qquad
\Delta_{\varphi} \equiv 1-2\eps/3,
\qquad
f(u) \equiv R(1,u,g_{*}).
\label{differ4}
\end{equation}
From Eq. (\ref{differ4}) it follows that in the IR range (large $\mu r$ and
arbitrary $mr$) the parameters $g_{0}$ and $\nu_0$ enter into the
correlation functions only in the form of the combination
$D_{0}=g_{0}\nu_0^{3}$, a fact first established in Ref. \cite{Frisch}.
For $\eps=2$ this proves the Second Kolmogorov hypothesis: owing to the
relation (\ref{2.74}), the correlation functions depend on $\E$ but do not
depend on the viscosity coefficient $\nu_0$ and the UV scale
$\Lambda\sim g_{0}^{1/2\eps}$; see also the discussion in \cite{turbo,JETP}.

Representation (\ref{differ4}) for any scaling function $f(u)$ describes the
behavior of the correlation functions for $s\equiv \mu r \gg1$ and any fixed
value of $u\equiv mr$; the inertial range corresponds to the additional
condition $u\ll1$. The form of the function $f(mr)$ is not determined by the
RG equation (\ref{RGE}). Calculating the function $R$ in (\ref{dimm}) within
the renormalized perturbation theory, $R=\sum_{n=0}^{\infty} g^{n}R_{n}$,
substituting $g\to g_{*}$ and expanding $g_{*}$ and $R_{n}$ in $\eps$, one
obtains the $\eps$ expansion for the scaling function:
\begin{equation}
f(u)=\sum_{n=0}^{\infty}\eps^{n}f_{n}(u).
\label{1.64}
\end{equation}

Although the coefficients $f_{n}$ in our model are finite at $u=0$, this
does not prove the finiteness of $f(u)$ beyond the $\eps$ expansion: one
can show that for any arbitrarily small value of $\eps$ there are diagrams
that diverge at $m\propto u \to 0$. As a result, the coefficients
$f_{n}$ contain IR singularities of the form $u^{p} \ln^{q}u$, these
``large IR logarithms'' compensate for the smallness of $\eps$, and the
actual expansion parameter appears to be $\eps\ln u$ rather than $\eps$
itself. Thus the plain expansion (\ref{1.64}) is not suitable for the
analysis of the small-$u$ behavior of $f(u)$.

The formal statement of the problem is to sum up the expansion (\ref{1.64})
assuming that $\eps$ is small with the additional condition that
$\eps\ln u\sim1$. By analogy with the theory of critical behavior
\cite{Zinn,book}, the desired solution can be obtained using the well-known
operator product expansion (OPE); see \cite{UFN,turbo,JETP}. The OPE
shows that for small $u$ the function $f(u)$ has the form
\begin{equation}
f(u)= \sum_{F} C_{F}(u)\, u^{\Delta_{F}},
\label{SDE}
\end{equation}
where $C_{F}$ are coefficients analytical in $u^{2}$, the summation runs over
all local composite operators $F(x)$ consistent with the symmetries of the
model and the left-hand side, and $\Delta_{F}$ are their scaling dimensions
calculated as series in $\eps$.

If all $\Delta_{F}$ are non-negative (as in models of critical behavior),
expression (\ref{SDE}) is finite at $u=0$ and the $\eps$ expansion can be
used for the calculation of $f(0)$.
The feature specific to statistical models of fluid turbulence is the
existence of composite operators with {\it negative} dimensions $\Delta_{F}$.
Such operators, termed ``dangerous'' in \cite{UFN,turbo,JETP}, give rise
to strong IR singularities in the correlation functions and to their
divergence at $m\propto u\to0$.

The problem is that dangerous operators in our model are absent in the
$\varepsilon$ expansions and can appear only at finite values of
$\varepsilon$. This means that they can be reliably identified only if
their dimensions are derived {\it exactly} with the aid of Schwinger
equations or Ward identities that express Galilean symmetry. Due to the
nonlinear nature of the problem, they enter the corresponding OPE's as
infinite families whose spectra of dimensions are not bounded from below,
and in order to find the small-$mr$ behavior one has to sum up all their
contributions in the representation (\ref{SDE}). The needed summation of
the most singular contributions, related to the operators of the form
$\varphi^{n}$ with known dimensions, was performed in \cite{JETP}
using the so-called infrared perturbation theory for the case of the
different-time pair correlation functions; see also \cite{UFN,turbo}.
It has revealed their strong dependence on $m$, which physically can
be explained by the infamous ``sweeping effects.'' This demonstrates that,
contrary to the existing opinion \cite{K,CK,Teo,Teo2,Woo}, the sweeping
effects can be properly described within the RG approach, but one should
combine the RG with the OPE and go beyond the plain $\eps$ expansions.

In Secs. \ref{sec:pair}--\ref{sec:CK}, we shall be interested in the
Galilean invariant objects like the equal-time structure functions. For such
objects, the singular contributions mentioned above drop out from the
representation (\ref{SDE}) in agreement with the fact that Galilean
invariant quantities are not affected by the sweeping. Only the dimensions
of Galilean invariant operators can appear in (\ref{SDE}), and the singular
behavior can be related to invariant operators with $\Delta_{F}<0$.
No such operator, however, has been presented in any study we know of;
see Refs. \cite{Pismak,Kim,Kim2}. Although the problem should be considered
open, this fact suggests that Galilean invariant objects may have a finite
limit at $u\to0$. Below in Sec.~\ref{sec:CK} we shall calculate the
Kolmogorov constant within the assumption that the second-order structure
function has a finite limit at $u\to0$ for the physical value $\eps=2$.
Such an assumption can be justified by the real experiments, which
indicate that the corresponding exponent is hardly distinguishable from
the Kolmogorov value \cite{Monin}, or by the example of the Kraichnan
model, in which the second-order function is not anomalous \cite{Kraich1}.

\section{Two-loop calculation of the renormalization constant,
RG functions, fixed point and the UV correction exponent}
\label{sec:ZZ}

Let us turn to the calculation of the renormalization constant $Z_{\nu}$
in Eq. (\ref{1.30}) with the accuracy $O(g^{2})$ (two-loop approximation).

The renormalization constants in the MS scheme are independent of the value
of the ``mass''  $m$ and the specific form of the function $h(m/k)$ in the
correlator (\ref{1.9}), and the practical calculations are usually performed
in the ``massless'' model (\ref{1.10}) with $m=0$ and $h(m/k)=1$.
It turns out, however, that for the multiloop calculations it is more convenient
to use the
model (\ref{1.9}) with $m\ne0$ and $h(m/k)=\theta(k-m)$.
We shall determine the constant $Z_{\nu}$ from the
requirement that the 1-irreducible correlation function
$\langle\varphi'\varphi\rangle$ at zero frequency ($\omega=0$) and in the
limit ${p}\to0$ be UV finite (that is, be finite at $\eps\to0$) when
expressed in renormalized variables using relations (\ref{18}). More
precisely, we shall consider the scalar dimensionless quantity
\begin{eqnarray}
\Gamma= \lim_{{ p}\to 0}
\frac{\bigl\langle\varphi'_{i}\varphi_{i}\bigr\rangle_{{\rm
1-ir}}\, ({\bf p}; \omega=0)} {\nu p^2(d-1)},
\label{Ratio}
\end{eqnarray}
which depends only on $g$ and $m/\mu$ and in the loopless (tree)
approximation equals $\Gamma^{(0)}=-Z_{\nu}$. The IR regularization
is provided by the sharp cutoff  $h(m/k)=\theta(k-m)$
in Eq. (\ref{1.9}).

Let us illustrate this scheme by revisiting the one-loop calculation.
To first order in $g$, the quantity (\ref{Ratio}) is determined by the
only diagram
\begin{equation}
\epsfig{file=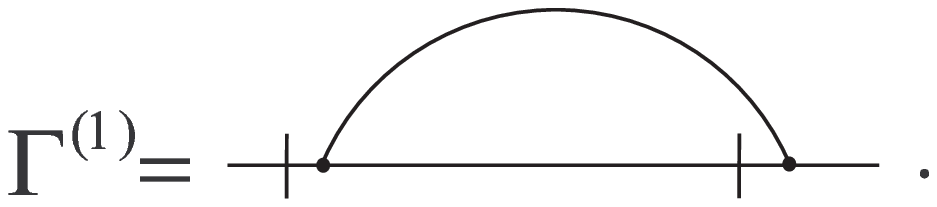,width=4.5cm}
\label{Diagram}
\end{equation}
Here the propagators (\ref{lines}) are represented by the lines, the slashed
ends correspond to the field $\varphi'$, the ends without a slash correspond
to $\varphi$. The triple vertices correspond to the vertex factor $V_{ijs}$ (see the text below Eq. (3.3)),
and the contraction with the transverse projector is implied. Integrating
over the frequency and taking the limit ${ p}\to 0$ gives
\begin{eqnarray}
\Gamma^{(1)}=\,-\,\frac{g\mu^{2\eps}Z_{\nu}^{-2}}
{4(d-1)(2\pi)^{d}}\int\frac{d {\bf k}}{k^{d+2\eps}}\,
 h(m/k) \bigl[d-3+(9-d) z^{2}-6z^{4}\bigr],
\label{A11}
\end{eqnarray}
where $z\equiv({\n\cdot\k})/{k}$ is the cosine between ${\bf k}$ and
$\n\equiv{\p}/{p}$, the direction of the external momentum.
The quantity $\Gamma$ does not depend on the direction $\n$ and we
can average (\ref{A11}) with respect to it using the relations
\begin{eqnarray}
\langle z^{2n}\rangle=\frac{(2n-1)!!}{d(d+2)\dots(d+2n-2)},
\qquad \langle z^{2n+1}\rangle=0;
\label{A12}
\end{eqnarray}
the brackets denote the averaging over the unit sphere in $d$ dimensions.
This gives
\begin{eqnarray}
\Gamma^{(1)}=-\frac{g\mu^{2\eps}\, \bar S_{d}\,(d-1)}
{4(d+2)Z_{\nu}^{2}} \,
\int_{m}^{\infty}\frac{dk}{k^{1+2\eps}}=
-\frac{(d-1)(\mu/m)^{2\eps}\, g\bar S_{d}}{8(d+2)\,\eps\, Z_{\nu}^{2}}
\label{A13}
\end{eqnarray}
with $\bar S_{d}$ from (\ref{surface}).
In the expansion in the number of loops,
\begin{eqnarray}
\Gamma=-Z_{\nu}+\Gamma^{(1)}+\Gamma^{(2)}+\dots,
\label{A14}
\end{eqnarray}
the $n$-loop term $\Gamma^{(n)}$ is the quantity of order $O(g^{n})$, so that
in the $O(g)$ approximation in (\ref{A14}) it is sufficient to keep the term
$\Gamma^{(1)}$ (and put $Z_{\nu}=1$ in it) and the linear in $g$ contribution
in the first term. The residue at the pole in $\eps$ in (\ref{A13}) is found
by the simple substitution $(\mu/m)^{2\eps}\to1$. From the requirement that
the function $\Gamma(g,m/\mu)$ be UV finite (that is, finite at $\eps\to0$)
we thus find $a_{11}=-(d-1)\bar S_d/8(d+2)$ for the coefficient $a_{11}$ in
agreement with the well-known result (\ref{a11}).

The two-loop contribution $\Gamma^{(2)}$ in (\ref{A14}) is given by the sum
of the following eight diagrams (normalization according to (\ref{Ratio})
is implied):
\begin{equation}
\epsfig{file=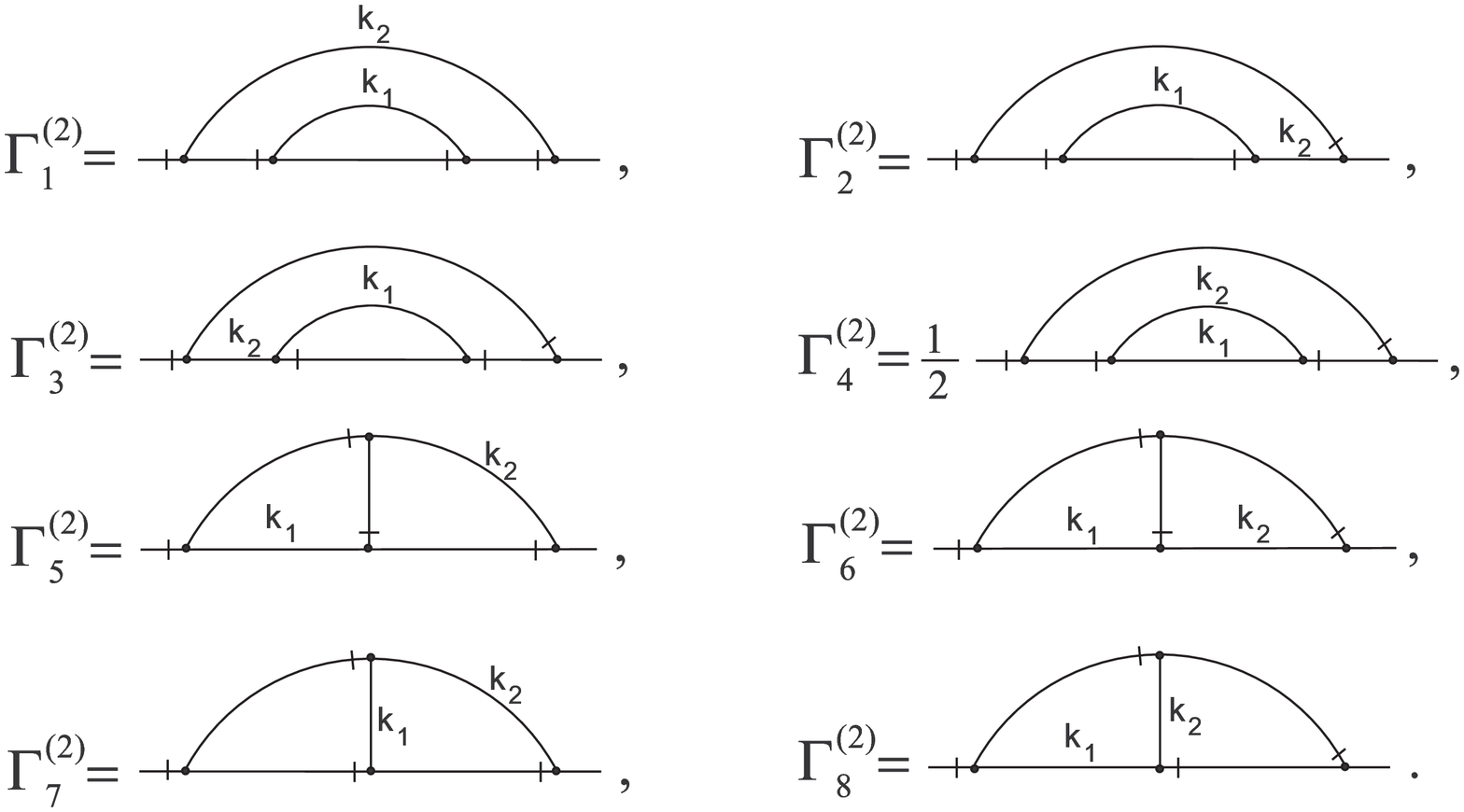,width=12cm}
\label{A16}
\end{equation}

After two simple integrations over the frequencies, these diagrams are
represented as integrals over two independent momenta ${\bf k}_{1}$,
${\bf k}_{2}$ that can be assigned to the two
$\langle\varphi\varphi\rangle_{0}$ lines, while the external momentum
$\p$ flows only via the $\langle\varphi\varphi'\rangle_{0}$ lines
(such choice of the integration contours is possible for any of the
diagrams in (\ref{A16})). The way in which the momenta ${\bf k}_{1}$
and ${\bf k}_{2}$ are assigned to the $\langle\varphi\varphi\rangle_{0}$
lines is explicitly shown in (\ref{A16}); additional symmetrization in
${\bf k}_{1}$, ${\bf k}_{2}$ is understood for the diagrams
$\Gamma^{(2)}_{i}$ with $i=5,\dots,8$. The diagram
$\Gamma^{(2)}_{4}$ is automatically symmetric.

After the contractions of the tensor indices
and the limit transition $p\to0$ have been performed, the resulting
integrands take on the form of polynomials in the cosines
$z_{1}=(\n\cdot {\bf k}_{1})/k_{1}$ and
$z_{2}=(\n\cdot {\bf k}_{2})/k_{2}$. The integrands are further
simplified after averaging over the direction $\n\equiv\p/p$ using
(\ref{A12}), which is possible owing to the independence of the quantities
$\Gamma_{i}^{(2)}$ of $\n$. In the variables ${\bf k}_{1}$, ${\bf k}_{2}$
and $z\equiv({\bf k}_{1}\cdot{\bf k}_{2})/k_{1}k_{2}$ they can be written
in the form
\begin{equation}
\Gamma_{i}^{(2)}=g^{2} \bar S_{d}^{2}\, \mu^{4\eps}\, Z_{\nu}^{-5}\,
\int_{m}^{\infty} \frac{dk_{1}} {k_{1}^{1+2\eps}}
\int_{m}^{\infty} \frac{dk_{2}} {k_{2}^{1+2\eps}}
\int_{0}^{1}dzf_{i}(z,k_{1}/k_{2}).
\label{A17}
\end{equation}
The lower limits in the integrals over $k_{1}$, $k_{2}$ are due to the
choice $h(m/k)=\theta(k-m)$. For convenience reasons, here and in analogous
formulas below we restrict the integration over $z$ to the half-interval
$[0,1]$ instead of the full interval $[-1,1]$ and simultaneously replace
the integrands by their doubled even-in-$z$ parts. The expressions for the functions ${f}_{i}$ are given in the Appendix.
Within our accuracy, it is necessary to replace $Z_{\nu}\to1$ in (\ref{A17}).

In terms of the dimensionless variables $\kappa_{i}\equiv k_{i}/m$
one can write
\begin{equation}
\Gamma_{i}^{(2)}=g^{2} \bar S_{d}^{2} (\mu/m)^{4\eps}A_{i}(\eps)
\label{A18}
\end{equation}
with
\begin{equation}
A_{i}(\eps)=\int_{1}^{\infty} \frac{d\kappa_{1}}{\kappa_{1}^{1+2\eps}}
\int_{1}^{\infty} \frac{d\kappa_{2}}{\kappa_{2}^{1+2\eps}}
\int_{0}^{1}dz f_{i}(z,\kappa_{1}/\kappa_{2}).
\label{A19}
\end{equation}
We are interested in the coefficients $a_{i}$, $b_{i}$ in the pole part
of the quantities $A_{i}(\eps)$:
\begin{equation}
A_{i}(\eps)=\frac{a_{i}}{\eps^{2}}+\frac{b_{i}}{\eps}+c_{i}+O(\eps).
\label{A20}
\end{equation}

The subdiagram in $\Gamma_{4}^{(2)}$ does not contain an UV divergence,
so that $f_{4}(z,0)=f_{4}(z,\infty)=0$, the integrals over $k_{1}$ and
$k_{2}$ are separately finite, and the divergence at $\eps\to0$ comes only
from the region where $k_{1}$, $k_{2}$ tend to infinity simultaneously.
As a result, the second-order pole is absent: $a_{4}=0$. The subdiagrams in
$\Gamma_{5}^{(2)},\dots,\Gamma_{8}^{(2)}$ contain UV divergences, but in
the sum $\Gamma_{0}^{(2)}\equiv\sum_{i=5}^{8}\Gamma_{i}^{(2)}$ they cancel
each other (a consequence of the absence of divergence in the 1-irreducible
function $\langle\varphi'\varphi\varphi\rangle$), so that
$a_{0}\equiv\sum_{i=5}^{8} a_{i}=0$.

For the diagrams $\Gamma_{i}$ with $i=1,2,3$ one still has $f_{i}(z,0)=0$,
but the second-order poles exist due to the relations $f_{i}(z,\infty)={\rm const}\ne0$.
Expressions (\ref{A19})  can be simplified using the
identity
\begin{equation}
{\cal D}_{m}\Gamma_{i}^{(2)}=-4\eps g^{2} \bar S_{d}^{2} (\mu/m)^{4\eps}
\, A_{i}(\eps)
\label{mpartial}
\end{equation}
that follows from (\ref{A18}). The operation $ {\cal D}_{m}
\equiv m\partial/\partial m$ reduces the number of integrations in
(\ref{A19}) and explicitly isolates the pole factor $\eps^{-1}$:
calculating the left-hand side of (\ref{mpartial}) using (\ref{A17})
and changing to dimensionless variables again, one obtains
\begin{equation}
A_{i}(\eps)=\frac{1}{4\eps}\int_{1}^{\infty}
\frac{d\kappa}{\kappa^{1+2\eps}} \int_{0}^{1}dz
\left[ f_{i}(z,\kappa)+f_{i}(z,1/\kappa) \right].
\label{A21}
\end{equation}

For $i=0,4$ the integral in (\ref{A21}) is finite at $\eps=0$ and determines
the residue at the first-order pole:
\begin{equation}
a_{i}=0,\qquad
b_{i}=\frac{1}{4}\int_{1}^{\infty} \frac{d\kappa}{\kappa}
\int_{0}^{1}dz
\left[f_{i}(z,\kappa)+f_{i}(z,1/\kappa)\right], \quad i=0,4.
\label{A22}
\end{equation}

For the quantities $A_{i}(\eps)$ with $i=1,2,3$ the residue at the
second-order pole is obtained if the value of the function
$\left[f_{i}(z,\kappa)+f_{i}(z,1/\kappa)\right]$ in (\ref{A21}) is
substituted by its limit value $f_{i}(z,\infty)+f_{i}(z,0)=f_{i}(z,\infty)$
[we recall that $f_{i}(z,0)=0$];
then the integration over $\kappa$ becomes simple and gives
\begin{equation}
a_{i}=\frac{1}{8}\int_{0}^{1}dzf_{i}(z,\infty), \quad i=1,2,3.
\label{A23}
\end{equation}
The remaining integral with the replacement $f_{i}(z,\kappa)\to
\left[f_{i}(z,\kappa)-f_{i}(z,\infty)\right]$ is finite at $\eps=0$ and
determines the residue at the first-order pole:
\begin{equation}
b_{i}=\frac{1}{4}\int_{1}^{\infty} \frac{d\kappa}{\kappa} \int_{0}^{1}dz
\left[f_{i}(z,\kappa)-f_{i}(z,\infty)+f_{i}(z,1/\kappa)\right],
\quad i=1,2,3.
\label{A24}
\end{equation}

According to (\ref{A18}), (\ref{A20}), the summed values
\begin{equation}
a\equiv\sum_{i=0}^{4}a_i, \quad b\equiv\sum_{i=0}^{4}b_i
\label{b}
\end{equation}
of the coefficients
$a_i$, $b_i$ from (\ref{A22})--(\ref{A24}) determine the needed residues
for $\Gamma^{(2)}=\sum_{i=0}^{4}\Gamma^{(2)}_i$. Analysis shows that all
integrations over $z$ in (\ref{A22})--(\ref{A24}) can be performed in terms
of elementary functions for any $d$, which gives simple expressions for $a_{i}$.
For their summed value $a$ in (\ref{b}) this gives:
\begin{equation}
a = \frac{(d-1)^2}{64(d+2)^2} = \bar S_{d}^{-2}\, a_{11}^{2}
\label{a}
\end{equation}
with $\bar S_{d}$ from Eq. (\ref{surface}) and $a_{11}$ from (\ref{a11}).

It turns out, however, that the most convenient way to find the residues $b_{i}$
is the direct numerical calculation of the double integrals in (\ref{A22}),
(\ref{A24}), which can be performed for any given value of $d$.
Below we give the results of such calculation for the most interesting
case $d=3$; it is convenient to represent them in terms of
the quantities  $b'_{i}\equiv 10^3\cdot b_{i}$ and
$b'\equiv\, 10^3\cdot b = \sum_{i=0}^{4} b'_{i}$:
\begin{equation}
b'_{0}=-1.34, \quad b'_{1}=0.16, \quad
b'_{2}=1.21, \quad b'_{3}=0.006, \quad
b'_{4}=-4.17, \quad b'=-4.13.
\label{bb}
\end{equation}

Now let us write down the condition of the UV finiteness of the
right-hand side of Eq. (\ref{A14}) in the second order in $g$. In
Eq. (\ref{A17}) it was sufficient to replace $Z_{\nu}\to1$, while
in the expression (\ref{A13}) for $\Gamma^{(1)}$ one should keep
in the expansion $Z_{\nu}^{-2}=1-2a_{11}g/\eps+\dots$ the term
linear in $g$. Then using Eq. (\ref{1.30}) the condition of the UV
finiteness is written in the form
\begin{equation}
-\left( a_{22}\eps^{-2}+a_{21}\eps^{-1} \right) - 2a_{11}^{2}\eps^{-2}
\left( 1+2\eps\ln\frac{\mu}{m} \right) +
\bar S_{d}^{2}a\eps^{-2}\left( 1+4\eps\ln\frac{\mu}{m} \right) +
\bar S_{d}^{2}b\eps^{-1}=0.
\label{A25}
\end{equation}
From Eq. (\ref{a}) it is easily seen that the terms with $\ln(\mu/m)$
in (\ref{A25}) cancel each other (a consequence of the renormalizability
of the model in the MS scheme). For the sought
coefficients $a_{22}$ and $a_{21}$ in (\ref{1.30}) from Eqs. (\ref{A25})
and (\ref{a})  we thus obtain
\begin{equation}
a_{22} =  -a_{11}^2,
 \qquad a_{21} = \bar S_{d}^2\, b
\label{A26}
\end{equation}
with  $a_{11}$ from Eq. (\ref{a11}) and $b$ from (\ref{b}).
This completes the calculation of the renormalization constant $Z_{\nu}$
in the two-loop approximation.

The knowledge of the renormalization constant $Z_{\nu}$ to order $O(g^{2})$
allows for the calculation of the anomalous dimension $\gamma_{\nu}$ in
(\ref{RGF1}) to order $O(g^2)$:
\begin{equation}
\gamma_\nu(g) =
-2g\partial_g a_1(g)=-2\left(a_{11}g+2a_{21}g^2\right) +O(g^3),
\label{RGF}
\end{equation}
then the $\beta$ function is found from (\ref{RGF1}) with the proper
accuracy. The coordinate of the fixed point in (\ref{FP1}) is then
determined to order $O(\eps^2)$. From Eqs. (\ref{A26}) and (\ref{RGF})
we obtain:
\begin{equation}
g_{*} \bar S_{d} = \alpha \eps  (1+\lambda\eps) + O(\eps^{3}),
\quad
\alpha \equiv - \bar S_{d}/3a_{11} = 8(d+2)/3(d-1),
\quad
\lambda\equiv 2a_{21}/3a_{11}^{2} = 2b [8(d+2)/(d-1)]^{2}/3
\label{FPd}
\end{equation}
with the coefficient $b$ from (\ref{b}) and $\bar S_{d}$ from
(\ref{surface}). In particular, for $d=3$ using Eq. (\ref{bb}) we obtain:
\begin{equation}
g_{*} = (40 \pi^2\eps/3) (1+\lambda\eps) + O(\eps^{3}),
\quad \lambda  \simeq -1.101.
\label{FP}
\end{equation}
The case of general $d$ dimensions is also of interest and will be
discussed in Sec.~\ref{sec:CK}. Results for the coefficient $\lambda$
for different values of $d$ are given in Table~\ref{table2} along with
other quantities that will appear later on.

The correction exponent $\omega=\beta'(g_{*})$ in (\ref{FP1}), determined
by the slope of the $\beta$ function at the fixed point, is found from
(\ref{RGF}) and (\ref{FP1}) to second order of the $\eps$ expansion:
\begin{equation}
\omega=2\varepsilon(1-\lambda\varepsilon)+O(\varepsilon^3).
\label{Omega}
\end{equation}
From Table~\ref{table2} it follows that, for any $d$, the two-loop correction
makes the value of $\omega$ ``more positive," that is, it enhances the IR
stability of the fixed point.

\section{Pair correlation function: $\varepsilon$ expansion} \label{sec:pair}

In this section we shall discuss the equal-time pair correlation function
of the velocity field in the momentum representation,
\begin{equation}
G_{ij}({\bf p})=P_{ij}({\bf p})\, G(p).
\label{G}
\end{equation}
The analogs of representations (\ref{dimm}) and (\ref{differ4})
for $G(p)$ have the forms:
\begin{equation}
G(p) =  g\nu^2p^{-d+2} R(s,u,g) \simeq D_{0}^{2/3}g_{*}^{1/3}
p^{-d+2\Delta_{\varphi}} R(1,u,g_{*}) , \qquad s\equiv p/\mu,
\quad u\equiv m/p,
\label{dimG}
\end{equation}
where the second relation holds in the IR asymptotic region and one factor
$g$ is explicitly isolated such that the expansion in $g$ of the
dimensionless function $R(1,u,g)$ begins with $O(g^{0})$.

Our aim is twofold. Below in this section we shall calculate two orders
(1 and $\eps$)
of the $\eps$ expansion for the scaling function $R(1,u,g_{*})$ at $u=0$
(this accuracy is consistent with the two-loop calculation of the RG
functions in Sec.~\ref{sec:ZZ}). The coefficients of the $\eps$ expansions
in our model have a finite limit at $u=0$ (see the end of
Sec.~\ref{sec:QFT}) so we can perform this calculation in the
``massless'' model (\ref{1.10}) with $m=0$ and $h(m/k)=1$, which is
always assumed below unless stated to be otherwise. These results will
be used later in Sec.~\ref{sec:CK} in the calculation of the skewness
factor and the Kolmogorov constant.
In Sec.~\ref{sec:bsk}, we shall use the example of the pair correlator
to discuss the existence of the limit $u\to0$ for finite $\eps$, the
possibility of the extrapolation of the $\eps$ expansion beyond the
threshold where the sweeping effects become important (see the remark
in the end of Sec.~\ref{sec:QFT}), and the relation between
the RG approach and other approaches to the stochastic NS equation.

In order to calculate two terms of the $\eps$ expansion of the amplitude
$R(1,0,g_{*})$ we need to find the coefficients $c_{1}$ and $c_{2}$ in
the expansion
\begin{equation}
R(1,0,g)=G(p)/ \left(g\nu^2p^{-d+2}\right)|_{p=\mu, u=0}=
c_{1} + g\bar S_{d} c_{2} +O(g^{2})
\label{35}
\end{equation}
(the factor $\bar S_{d}$ from Eq. (\ref{surface}) is isolated in the second
term for convenience). We shall see below that $c_{1}=1/2$, while the
coefficient $c_{2}=c_{2}(\eps)$ should be found to the leading order in
$\eps$, that is, at $\eps=0$.

With the needed accuracy of $O(g^2)$, the function $G(p)$ is given by the
sum of the loopless (tree) and one-loop diagrams. The first (tree)
contribution to the correlator (\ref{G}) is simply given by the
bare correlator $\langle\varphi\varphi\rangle _0$ from (\ref{lines2})
at $t=t'$ and expressed in renormalized variables using Eq. (\ref{18}).
The corresponding contribution $R_{0}$ to the dimensionless quantity
(\ref{35}) has the form
\begin{equation}
R_{0} =  Z_{\nu}^{-1}/2 = \left[1-a_{11}g/\eps+O(g^{2})\right]/2
\label{D0}
\end{equation}
with the coefficient $a_{11}$ from (\ref{a11}). This gives the exact
result $c_{1} = 1/2$ for the first coefficient in (\ref{35}).

The one-loop contribution to the function $G(p)$ is given by the
sum of the following diagrams
\begin{equation}
\epsfig{file=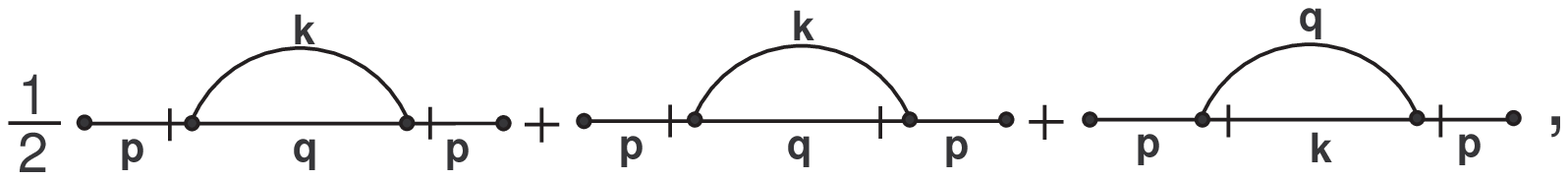,width=12cm}
\label{DiA}
\end{equation}
the diagrammatic notation was explained below Eq. (\ref{Diagram}).
We shall denote their contributions to the dimensionless function (\ref{35})
as $R_{1,2,3}$ (from the left to the right). They have the forms
\begin{mathletters}
\label{DDD}
\begin{eqnarray}
R_{1} &=& \frac{g}{8(d-1)}  \int \frac{d{\bfkap}}{(2\pi)^{d}}\,
U({\bfkap}, \q)\, Q_{1}({\bfkap}) \, \kappa^{2-d-2\eps} q^{2-d-2\eps},
\label{D1}  \\
R_{2} &=& \frac{g}{8(d-1)}  \int \frac{d{\bfkap}}{(2\pi)^{d}}\,
U({\bfkap}, \q)\, Q_{2}({\bfkap})\, \kappa^{2-d-2\eps} ,
\label{D2} \\
R_{3} &=& \frac{g}{8(d-1)}  \int \frac{d{\bfkap}}{(2\pi)^{d}}\,
U({\bfkap}, \q)\, Q_{2}({\bf q}) \, q^{2-d-2\eps} ,
\label{D3}
\end{eqnarray}
\end{mathletters}
where we included the symmetry coefficient $1/2$ for $R_{1}$ and denoted
\begin{eqnarray}
&Q&_{1}({\bfkap})= d-1 -2d ({\bf n}\cdot{\bfkap})+2(d-2)\kappa^{2}+4
({\bf n}\cdot{\bfkap})^{2}, \nonumber \\
&Q&_{2}({\bfkap}) =1-d +2(d-1)({\bf n}\cdot{\bfkap})-(d-3)\kappa^{2}
-2\kappa^{2}({\bf n}\cdot{\bfkap}), \nonumber \\
&U&({\bfkap}, \q) = [\kappa^{2}-({\bf n}\cdot{\bfkap})^{2}] /
\kappa^{2} q^{2} (1+\kappa^{2}+q^{2}).
\label{Q}
\end{eqnarray}
In Eqs. (\ref{DDD}) and (\ref{Q}) we use only dimensionless variables,
namely, the momenta divided by the modulus $p$ of the external momentum:
${\bf n}\equiv {\bf p}/p$ is the direction of the external momentum,
$\bfkap\equiv {\bf k}/p$ is the momentum flowing via the
$\langle\varphi\varphi'\rangle _0$ line for $R_3$, the
$\langle\varphi\varphi\rangle _0$ line for $R_2$ and
any one of the two equivalent
$\langle\varphi\varphi\rangle _0$ lines for $R_{1}$, and
${\bf q} = {\bf n} - {\bfkap}$ is the momentum flowing via the
remaining line. In order to isolate the scalar factor $G(p)$ we
contracted the full expression (\ref{G}) with the transverse
projector $P_{ij}({\bf p})$  and divided the result by its trace;
this explains the origin of the factor $P_{ii}({\bf p})=(d-1)$
in the denominator.

It is easily checked that $Q_{1}({\bfkap})=Q_{1}({\q})$ and
$\kappa^{2} - ({\bf n}\cdot{\bfkap})^{2}= q^{2} - ({\bf n}\cdot{\q})^{2}$.
From the last relation it follows that $U({\bfkap}, \q)=U(\q, {\bfkap})$
and therefore $R_{2}=R_{3}$. Another consequence of the last relation is
that $U({\bfkap}, \q)$ remains finite at $\bfkap\to0$ and $\q\to0$,
so that the IR convergence of the integrals (\ref{DDD})
at $\bfkap\to0$ and $\q\to0$ are determined by the factors
$\kappa^{2-d-2\eps}$ and $q^{2-d-2\eps}$, respectively. The IR
divergences in all three integrals (\ref{DDD}) arise for $\eps\ge1$,
but in their sum they cancel each other: the singularities at
$\bfkap=0$ (when $q^{2}= 1$)
in the integrands (\ref{D1}) and (\ref{D2}) cancel
each other due to the relation
\[ Q_{1}({\bfkap})|_{\kappa=0} = (d-1) = - Q_{2}({\bfkap})|_{\kappa=0}, \]
while the singularities at $\q=0$ (when $\kappa^{2}=1$)
in the integrands (\ref{D1}) and (\ref{D3}) cancel out due to the relation
\[ Q_{1}({\bfkap})|_{q=0}=Q_{1}({\q})|_{q=0}=(d-1)=-Q_{2}({\q})|_{q=0}, \]
where we have used the identity $Q_{1}({\bfkap})=Q_{1}({\q})$.
Thus the IR divergences that appear in the individual diagrams $R_{1,2,3}$
at $\eps=1$ cancel each other in their sum, and the latter becomes IR
divergent only for $\eps=2$. These general facts will be illustrated by
explicit formulas for the limit $d\to\infty$ in Sec.~\ref{sec:bsk}.

Formulas (\ref{DDD}) are written for the ``massless'' model (\ref{1.10})
that is, $h=1$ in Eq. (\ref{1.9}). In the ``massive'' model with $h\ne1$
additional factors $h(u/\kappa)h(u/q)$, $h(u/\kappa)h(u)$, and
$h(u/q)h(u)$ appear in the integrands (\ref{D1}), (\ref{D2}), and
(\ref{D3}), respectively. Then the divergence of the integrals (\ref{DDD})
for $\eps\ge1$ is manifested in the form of the contributions $m^{2-2\eps}$,
divergent as $m\to0$ for $\eps\ge1$.

As the independent variables in the integrands in (\ref{DDD}) one can take
the modulus $\kappa$ and $z\equiv ({\n\cdot\bfkap})/\kappa$, the cosine
between the directions of the momenta ${\bf p}$ and ${\bf k}$. Then the
integrals over ${\bf k}$ in (\ref{DDD}) in the spherical coordinates
are written as
\begin{equation}
\int d{\bfkap} \dots\, = S_{d} \int_{0}^{\infty} d\kappa\, \kappa^{d-1}
\bigl\langle \dots \bigr\rangle =
S_{d-1} \int_{0}^{\infty} d\kappa\, \kappa^{d-1}
   \int_{-1}^{1} dz\, (1-z^{2})^{(d-3)/2} \dots
\label{VAR}
\end{equation}
with $S_{d}$ from Eq. (\ref{surface}). Like in Eq. (\ref{A12}),
the brackets denote the averaging over the unit sphere in $d$
dimensions, while the second relation is obtained by the integration
over all angles in the $(d-1)$-dimensional subspace orthogonal
to the vector ${\bf p}$.

Now we turn to the calculation of the coefficient
$c_{2}\equiv c_{2}|_{\eps=0}$ in (\ref{35}).
The quantity $R_{1}$ is UV finite and it is sufficient to calculate its
contribution at $\eps=0$. We put $\eps=0$ in (\ref{D1}) and
rewrite the integral in variables $z,\kappa$. This gives:
\begin{equation}
R_1|_{\eps=0}\equiv g \bar S_{d} {\cal D} =
\frac{g S_{d-1}}{16(d-1)(2\pi)^{d}}\, \int_0^\infty\, d\kappa\,
\int^1_{-1}dz\, (1-z^{2})^{(d-1)/2}\,\, \frac{\kappa \,
(d-1 -4 \kappa^2+ 2d \kappa^2-2dz\kappa+4z^{2}\kappa^2)}
{(1-2z\kappa+ \kappa^2)^{d/2} (1-z\kappa+ \kappa^2)}.
\label{A39}
\end{equation}

The integrals $R_2=R_3$ contain a pole in $\eps$
(manifestation of the UV divergence in the 1-irreducible function
$\langle\varphi'\varphi\rangle$; see Sec.~\ref{sec:QFT}) and
can be written as
\begin{eqnarray}
R_{2}=R_{3}=g \bar S_{d} \bigl[ {\cal A}/\eps+ {\cal B}+O(\eps)\bigr].
\label{400}
\end{eqnarray}
Expression (\ref{D2}) can be written in the form
\begin{eqnarray}
R_2 = \frac{g S_{d-1}}{16(d-1)(2\pi)^{d}}\,
\int_0^\infty \frac{d\kappa}{\kappa^{1+2\eps}}
\int^1_{-1}dz\, (1-z^{2})^{(d-1)/2}\, \psi(z,\kappa),
\label{41}
\end{eqnarray}
where
\begin{eqnarray}
\psi(z,\kappa) \equiv \frac{ (1-d+3 \kappa^2 - d \kappa^2
+2dz\kappa -2z\kappa- 2z\kappa^3)\,\kappa^2}
{(1-2z\kappa+ \kappa^2)(1-z\kappa+ \kappa^2)}.
\label{psi}
\end{eqnarray}
The integration region over $\kappa$ can be split in two parts: [0,1]
and [1,$+\infty$); the pole in $\eps$ comes only from the second part.
The expansion of $\psi(z,\kappa)$ at large $\kappa$ has the form
$\psi(z,\kappa)=-2z\kappa + (3-d-6z^{2})+O(1/\kappa)$. The first term is
odd in $z$ and vanishes after the integration over $z$; the second term
completely determines the residue at the pole:
\begin{eqnarray}
{\cal A} = \frac{S_{d-1}}{32(d-1)S_{d}}\,
\int^1_{-1}dz\, (1-z^{2})^{(d-1)/2}\, (3-d-6z^{2}) =
\frac{1}{32(d-1)}\, \left\langle (1-z^{2})(3-d-6z^{2})
\right\rangle = - \frac{(d-1)}{32(d+2)},
\label{residue}
\end{eqnarray}
where the brackets denote the averaging over the unit sphere in $d$
dimensions and Eqs. (\ref{A12}) and (\ref{VAR})
have been used. One can see that in
representation (\ref{35}), the total pole part $2{\cal A}/\eps$
of the diagrams $R_{2,3}$ and the pole that comes from expression
(\ref{D0}) cancel each other (a consequence of the renormalizability).
Thus the total contribution of the diagrams $R_{1,2,3}$ into Eq. (\ref{35})
has the form
\begin{eqnarray}
c_{2}\equiv c_{2}|_{\eps=0}=2{\cal B}+{\cal D}
\label{BD}
\end{eqnarray}
with ${\cal B}$ from (\ref{400}) and ${\cal D}$ from (\ref{A39}),
and for the first two terms of the $\eps$ expansion of the amplitude
$R(1,0,g_{*})$ using (\ref{FP1}) and (\ref{35}) we obtain
\begin{eqnarray}
R(1,0,g_{*}) = (1/2) \left[1+ 2\alpha c_{2}\, \eps  +O(\eps^{2})\right],
\label{epscal}
\end{eqnarray}
with $\alpha=8(d+2)/3(d-1)$ from (\ref{FP1}).

The quantity ${\cal B}$ can be found as follows. The integrals that remain
in (\ref{41}) after the pole part has been subtracted (that is, the integral
over [0,1] and the integral over [1,$+\infty$) with the substitution
$\psi(z,\kappa)\to\psi(z,\kappa)+2z\kappa- (3-d-6z^{2})$ in the integrand)
are UV finite and we may set $\eps=0$ in them. This gives:
\begin{eqnarray}
\frac{16(d-1) S_{d}}{S_{d-1}}
\,{\cal B}&=& \int_0^1 \frac{dk}{k} \int_{-1}^1 dz\,
(1-z^{2})^{(d-1)/2}\, \psi(z,\kappa)+
\int_1^\infty \frac{dk}{k}\int_{-1}^1 dz\, (1-z^{2})^{(d-1)/2}\,
\bigl[\psi(z,\kappa)+2z\kappa - (3-d-6z^{2}) \bigr]
\nonumber \\
&=&\int_1^\infty \frac{dk}{k} \int_{-1}^1 dz\, (1-z^{2})^{(d-1)/2}\,
\bigl[\psi(z,\kappa)+\psi(z,1/\kappa)+2z\kappa - (3-d-6z^{2})\bigr].
\label{42}
\end{eqnarray}

The integrals (\ref{A39}) and (\ref{42}) converge and can be evaluated
numerically for any given value of $d$. For $d=3$ one obtains:
\begin{equation}
{\cal B}=-0.000057, \quad {\cal D}=0.06699, \quad c_{2}= 0.0669.
\label{A44}
\end{equation}

The quantities (\ref{A44}) for other values of $d$ are given in
Table~\ref{table2}; we shall discuss them later in Sec.~\ref{sec:CK}.

\section{Finite $\varepsilon$: nonlocal interactions, infrared
singularities, and Galilean symmetry} \label{sec:bsk}

In a number of studies, the results of the RG approach to model (\ref{1.1}),
(\ref{1.2}), (\ref{1.9}) were interpreted and criticized in the language
traditional for the classical theory of turbulence, comparison with the
well-known direct interaction approximation (DIA) was made and the validity
of the $\eps$ expansion for finite $\eps\sim 1$ was called in question;
see e.g. \cite{K,CK,Teo,Teo2,Woo}.

Let us discuss the problem on the example of the pair correlation function
in the one-loop approximation (\ref{DiA}). The corresponding analytical
expressions (\ref{DDD}) involve three vectors $\p$, $\k$, $\q$ subject to
the restriction $\p=\k+\q$ (here, the momenta are {\it not} divided
by $p=|\p|$). In the following, the external momentum $\p$ will be taken
to lie in the inertial range. Then relevant contributions to (\ref{DDD}),
roughly speaking, can come from the three regions: $k\sim q \gg p$,
$p\sim q \gg k$, where ${\bf k}$ is the momentum flowing via the
$\langle\varphi\varphi\rangle _0$ line (any one of the two equivalent
$\langle\varphi\varphi\rangle _0$ lines for the diagram $R_{1}$), and
$p\sim k \sim q$ (all three momenta lie in the inertial range). We shall
refer to the corresponding contributions as being determined by
UV-nonlocal, IR-nonlocal, and local (in the momentum space) interactions,
respectively.

The typical argument against the use of the $\eps$ expansion for finite
$\eps$ can be formulated as follows. The diagrams $R_{2,3}$ have
poles in $\eps$ as a manifestation of the UV divergence at $\eps=0$. Thus
for small $\eps$ the leading contribution to those diagrams is determined
by UV-nonlocal interactions. For $\eps\ge\eps_{c}$ with certain
$\eps_{c}=O(1)$ the diagrams become IR divergent and do not exist without
the IR cutoff $m$ [$\eps_{c}=1$ for all diagrams $R_{1,2,3}$ in
(\ref{DiA})]. In the ``massless'' model (\ref{1.10}) with $m=0$ they
have the form of the poles in $\eps_{c}-\eps$.
Physically such IR divergences are explained by the so-called
sweeping effects, that is, the transport of small eddies as a whole by
large ones. This purely kinematic phenomenon has no effect on shaping
the energy flow over the spectrum and is irrelevant in determining
the exponents and amplitudes in power laws for Galilean invariant
quantities (e.g. equal-time structure functions). According to the
classical phenomenology the latter are formed by the interactions of
the eddies in the inertial range, that is, by local interactions
(in the above sense); see e.g. \cite{Legacy}.
One may thus believe that the extrapolation of the $\eps$ expansion
(reliable for $\eps\ll1$) to finite $\eps\sim1$ overestimates the
contribution of the UV-nonlocal interactions and misrepresents,
or neglects at all, the physically important contributions from the
IR-nonlocal and local interactions.

More specifically, the author of Refs. \cite{Woo} extended the
well-known direct-interaction approximation \cite{K59,Orszag,McComb}
to the problem (\ref{1.1}), (\ref{1.2}) with the power-law forcing
(\ref{1.10}). In the DIA, the pair
correlation function $\langle\varphi\varphi\rangle $ and the response
function $\langle\varphi\varphi'\rangle$ satisfy a closed system of
integro-differential equations; their right-hand sides involve
the skeleton analogs of the diagrams $R_{1,2,3}$ with exact (``dressed'')
lines and bare vertices. The author of Ref. \cite{Woo} used a scaling
Ansatz, expanded the integrals on the right-hand sides in $\eps$ and
retained their pole parts; this allowed him to reproduce some of the
results obtained earlier using the RG and the $\eps$ expansion.
On the other hand, one can expand the same integrals in the vicinity
of the pole in $\eps_{c}-\eps$ (for the {\it skeleton} one-loop diagrams
one has $\eps_{c}=3/2$) and obtain
a solution whose form is completely different from that of the solution
obtained within the $\eps$ expansion \cite{Woo}. Its extrapolation to
the physical value reproduces the well-known solution \cite{K59} of the
original DIA equations. Although the $\eps$ expansion fails to reproduce
this solution,  no definitive conclusion can be made at this point:
for the different-time pair correlation function, {\it neither} of the
aforementioned solutions is correct for $\eps\ge\eps_{c}$.

Indeed, it has long been realized that the DIA misrepresents the
contributions of the IR-nonlocal interactions, at least in the energy
spectrum
(it is determined by the equal-time pair correlation function and is
Galilean invariant); see Ref. \cite{K59,Orszag,McComb}.
Although the original Dyson--Wyld equations for the correlation and
response functions are exact, they involve infinite series of skeleton
diagrams; the practical solution implies a truncation (the one-loop
truncation gives the DIA) followed by the substitution of a scaling
Ansatz. As a result of the approximation, the Galilean symmetry is
violated, and the IR singularities, related to the sweeping effects,
do not cancel out in Galilean invariant quantities like the energy
spectrum.\footnote{The DIA is not a given-order approximation to the
exact Dyson--Wyld equations in any small, at least formal, expansion
parameter. It can be understood as the leading approximation for
$N\to\infty$ in certain extension of the original NS equation to the case
of $N$ interacting velocity fields \cite{WM}, but such an extension is
not Galilean invariant unless $N=1$.}
This leads to spurious strong dependence of the latter on
the IR scale and erroneous value (3/2 instead of 5/3) for the
corresponding exponent. No closed equation, however, can be written
for the {\it equal-time} correlation function, and the problem cannot
be solved by taking into account more diagrams.

The key difference between the RG and the self-consistency
approaches is that the former is based on the ordinary perturbation
theory, that is, expansion in the nonlinearity in (\ref{1.1}), which
involves a formal expansion parameter $g_{0}$. The problem (\ref{1.1}),
(\ref{1.2}), (\ref{1.9}) is Galilean invariant for any $g_{0}$, $\eps$
and $d$, so that the perturbation theory in $g_{0}$ is manifestly
Galilean invariant in any given order for any value of $\eps$ and $d$;
this is equally true for the renormalized perturbation theory (in $g$).
The use of the RG and OPE implies some infinite resummations of the
original perturbation series within controlled approximations and
therefore does not violate the Galilean symmetry of the original
perturbation expansion.

The plain $\eps$ expansion is indeed not suitable for the description
of the IR singularities related to the sweeping. Even in the standard
$\phi^{4}$ model of critical behavior, where the IR singularities are
weaker, $\eps$ expansions of the form (\ref{1.64}) cannot be used for
the analysis of the asymptotical behavior at $m\to0$. As already
discussed in the end of Sec.~\ref{sec:QFT}, this fact does not hinder
the use of the RG technique, which should be combined with the operator
product expansion to derive resummed representations of the form
(\ref{SDE}); see e.g. \cite{Zinn,book}. The distinguishing feature of our
model (\ref{1.1}), (\ref{1.2}), (\ref{1.9}) is the existence in
Eq. (\ref{SDE}) of negative dimensions $\Delta_{F}<0$. The summation of
the most singular contributions coming from the operators $\varphi^n$ was
performed in Ref. \cite{JETP}, the generalization to the case of a
time-dependent large-scale field is given in \cite{Kim2}. This
gives the adequate description of the sweeping effects within the RG
formalism; see also Refs. \cite{UFN,turbo}.

Admittedly,  negative dimensions can be reliably identified, and their
contributions can be summed up, only with the aid of nonperturbative
methods that allow one to go beyond the plain $\eps$ expansion: functional
Schwinger equations, Ward identities that express Galilean symmetry, and
infrared perturbation theory. These techniques are discussed in Refs.
\cite{UFN,turbo} in detail. At the moment, however, we are interested in
the equal-time pair correlation function (\ref{G}). As already mentioned
in the end of Sec.~\ref{sec:QFT}, the operators $\varphi^n$, as well as
other noninvariant operators, give no contribution to the corresponding
representation (\ref{SDE}) in agreement with the fact that Galilean
invariant quantities are not affected by the sweeping. Only the dimensions
of Galilean invariant operators can appear in (\ref{SDE}), and the singular
behavior can be related to invariant operators with $\Delta_{F}<0$.
No such operator, however, has been presented in any study we know of,
and for the time being we can assume that all dimensions $\Delta_{F}$
are non-negative. This fact suggests that the scaling function
$R(1,u,g_{*})$ (\ref{dimG}) in Sec.~\ref{sec:pair} is finite at $u=0$ and
the value of $R(1,0,g_{*})$ can be calculated within the framework of
the ``massless'' model (\ref{1.10}) and the $\eps$ expansion.

The cancellation of the IR divergences for $1<\eps<2$ in the sum of
diagrams (\ref{DDD}) was demonstrated in Sec.~\ref{sec:pair} for general
$d$. Below we consider them in the limit $d\to\infty$, where explicit
expressions for the integrals entering Eq. (\ref{DDD}) can be written for
all $0<\eps<2$ (in Sec.~\ref{sec:pair}, only two terms of the $\eps$
expansion were calculated).

In the diagrams $R_{2}=R_{3}$ nontrivial dependence on $d$ comes only
from the angular averaging. It follows from Eqs. (\ref{A12}) that for
$d\to\infty$, the angular averages behave as
$\langle z^{2n}\rangle\propto d^{-n}$. Thus in order to find the leading
term of the large-$d$ behavior, it is sufficient to discard all such
averages with $n\ge1$, or, equivalently, to put
$z= ({\bf n}\cdot{\bf k})/k = 0$ in the integrand (\ref{D2}). In the
variables $\kappa$, $z$
introduced above Eq. (\ref{VAR}) this leads to the replacements
$(1-z^{2}) \to 1$, $q^{2}\to (1+\kappa^{2})$,
$1+\kappa^{2}+q^{2} \to 2(1+\kappa^{2})$,
which along with the asymptotic relation
$Q_{2} =-d(1+\kappa^{2})+O(d^{0})$
for the quantity $Q_{2}$ in (\ref{Q}) gives:
\begin{equation}
R_{2}+R_{3}=2R_{2} \simeq \frac {-g}{8(2\pi)^d} \int d{\bfkap}\,\,
\frac{\kappa^{2-d-2\eps}} {(1+\kappa^{2})} = \frac {-g \bar S_{d}}{8}
\int^{\infty}_{0}\, d\kappa\,\,
\frac{\kappa^{1-2\eps}} {(1+\kappa^{2})} = \frac {-g \bar S_{d}}{16\eps}
\Gamma(1+\eps)\Gamma(1-\eps).
\label{inf1}
\end{equation}
The latter equality in (\ref{inf1}) holds for $\eps<1$; for $\eps\ge1$
the integral over $\kappa$ diverges at small $\kappa$.

In the integrand (\ref{D1}) for diagram $R_{1}$, nontrivial dependence
on $d$ comes also from the factor $q^{-d}$ and the procedure described
above cannot be applied. In the variables $\kappa$, $z$
expression (\ref{D1}) takes on the form
\begin{equation}
R_{1} = \frac {gS_{d-1}}{8(d-1)(2\pi)^d}
\int^{\infty}_{0} \,d\kappa\, \int^{+1}_{-1} dz\,
(1-z^{2})^{(d-1)/2}\,
\frac{Q_{1}\,\kappa^{1-2\eps}q^{-d-2\eps}}
{(1+\kappa^{2}+q^{2})}
\label{inf2}
\end{equation}
with $q^{2}=1+\kappa^{2}-2z\kappa$ and $Q_{1}$ from (\ref{Q}).
For large $d$ one can write
\[(1-z^{2})^{d/2}q^{-d} =(1-z^{2})^{d/2}(1+\kappa^{2}-2z\kappa)^{-d/2}=
\left\{ 1+ \frac{(\kappa-z)^{2}} {(1-z^{2})} \right\}^{-d/2} \simeq
\exp \left\{ - \frac{d}{2}\frac{(\kappa-z)^{2}} {(1-z^{2})} \right\}.\]
Along with the relation $S_{d-1}/S_{d} \simeq \sqrt{d/2\pi}\,
\left[1+O(1/d)\right]$, which follows from the Stirling formula for the
$\Gamma$ function that enters Eq. (\ref{surface}), this gives:
\[(1-z^{2})^{(d-1)/2}q^{-d} S_{d-1} \simeq S_{d} \delta(\kappa-z).\]
We substitute this relation into (\ref{inf2}), retain only intervals
$[0,1]$ in the integrals, use the relation $Q_{1}|_{k=z} =d+O(d^{0})$
for the quantity $Q_{1}$ in (\ref{Q}) and perform the integration over
any one of the variables; this gives:
\begin{equation}
R_{1} \simeq \frac {g\bar S_{d}}{8(d-1)}
\int^{1}_{0} d\kappa\, \int^{1}_{0} dz\,
Q_{1} \frac{\kappa^{1-2\eps}q^{-2\eps}} {(1+\kappa^{2}+q^{2})}
\delta(\kappa-z)
= \frac {g\bar S_{d}}{16} \int^{1}_{0} d\kappa\,
\frac{\kappa^{1-2\eps}} {(1-\kappa^{2})^{\eps}} =
\frac {-g \bar S_{d}}{32} \frac{\Gamma^{2}(1-\eps)}{\Gamma(2-2\eps)}.
\label{inf3}
\end{equation}
Like for the integral (\ref{inf1}), the latter equality holds for $\eps<1$;
for $\eps\ge1$ the integral over $\kappa$ diverges at small $\kappa\to0$
and $\kappa\to1$. Adding together the expressions (\ref{inf1}), (\ref{inf3})
with the loopless contribution (\ref{D0}), where the limit $d\to\infty$
of the explicit form (\ref{a11}) should be used for the coefficient $a_{11}$,
gives:
\begin{equation}
R(1,0,g) = \frac{1}{2}+ \frac {g \bar S_{d}}{32} {\cal R}(\eps), \quad
{\cal R}(\eps)\equiv\left\{\frac{2}{\eps}
  - \frac{2}{\eps} \Gamma(1-\eps)\Gamma(1+\eps) +
\frac{\Gamma^{2}(1-\eps)}{\Gamma(2-2\eps)}\right\}
\label{inf4}
\end{equation}
(the product $g\bar S_{d}$ at the fixed point has a finite limit for
$d=\infty$; see Eq. (\ref{FP1})).
The sum of the integrals (\ref{inf1}), (\ref{inf3}) converges for all
$0<\eps<2$: the divergences in the integrals (\ref{inf1}) and (\ref{inf3})
at $\kappa\to0$ cancel each other, while the cancellation of the divergence
in the integral (\ref{inf3}) at $\kappa\to1$ with the divergence in the
diagram $R_{3}$ becomes obvious after the replacement $\kappa\to1-\kappa$
in expression (\ref{inf1}) for $R_{2}$. Thus the expression (\ref{inf4})
for the {\it sum} of the integrals holds for the whole interval $0<\eps<2$.

In Eq. (\ref{inf4}), the cancellation of the poles at $\eps=0$ (a
consequence of the renormalizability) and $\eps=1$ (a consequence
of the Galilean symmetry) are obvious; the renormalizability and the
Galilean invariance of the model guarantee similar cancellations in
all the higher orders of the (renormalized) perturbation theory.

In models of quantum field theory, the elimination of UV divergences
is needed to ensure existence of meaningful (finite) results at the
physical value $\eps=0$. In our case it is not necessary to take
the limit $\eps\to0$ and ensure that it exists, that is, there is no
need to renormalize the model. However, it turns out that the possibility
itself of doing this implicitly contains valuable information about the
original model with finite $\eps$: it is this possibility that ensure
the validity of the RG equations (\ref{RGE}) and the operator product
expansion (\ref{SDE}).

Physically, elimination of the UV singularities (poles in $\eps$) removes
the contributions of the UV-nonlocal interactions that would obscure
relevant local interactions. Coefficients of the renormalized perturbation
series for the correlation functions have a finite limit at $\eps\to0$: they
are determined by the local interactions, while the UV-nonlocal interactions
are taken into account in the formulas of multiplicative renormalization
(\ref{18}). A fine correlation that exists between different orders of
the perturbation theory is expressed by the RG equations (\ref{RGE});
solving them allows one to perform certain infinite resummation of the
perturbation series for (renormalized) correlation functions.

Expression (\ref{inf4}) contains a pole at $\eps\to2$: $R(\eps) \simeq
6/(2-\eps)$ [for general $d$ we would obtain $R(\eps) \simeq 2(3d^{2}+d-12)
/ d(d+2)(2-\eps)$]; higher-order poles will necessarily appear in the
higher orders of the perturbation theory. Such singularities are necessary
to ensure the existence of the finite limit for the pair correlation
function at $\eps\to2$ in terms of the physical parameter $\E$ from
(\ref{1.3}). For the correlation function of $n$ fields $\varphi$, they
must combine into the singularity $(2-\eps)^{-n/3}$ that will cancel the
vanishing factor $D_{0}^{n/3} \propto \E^{n/3} (2-\eps)^{n/3}$ in
representation (\ref{differ4}); see Eq. (\ref{2.74}). The conjecture
that the singularities at $\eps\to2$ indeed combine into the needed
expressions is confirmed by two examples: the pair correlation function
in the exactly solvable Heisenberg model \cite{Hei} and the triple
correlation function in our case; see the exact expression (\ref{Exa})
in Sec.~\ref{sec:CK}.

\section{Two-loop calculation of the Kolmogorov constant} \label{sec:CK}

The Kolmogorov constant $C_{K}$ can be defined as the (dimensionless)
coefficient in the inertial-range expression $S_{2}(r)=C_{K}(\E r)^{2/3}$
for the second-order structure function, predicted by the Kolmogorov--Obukhov
theory and confirmed by experiment \cite{Monin,Legacy}. Here $\E$ is the mean
energy dissipation rate and the $n$-th order (longitudinal, equal-time)
structure function is defined as
\begin{equation}
S_{n} (r) \equiv \big\langle [ \varphi_{r} (t,{\bf x}+{\bf r})
- \varphi_{r} (t,{\bf x})]^{n} \big\rangle, \qquad
\varphi_{r}\equiv (\varphi_{i}\cdot r_{i})/r, \quad r\equiv |{\bf r}|.
\label{struc}
\end{equation}

Alternatively, the Kolmogorov constant $C_{K}'$ can be introduced via
the relation $E(k)=C_{K}' \E^{2/3} k^{-5/3}$, where the energy
spectrum $E(k)$ is related to the equal-time correlation function (\ref{G})
as $E(k)= \bar S_{d}(d-1)k^{d-1}G(k)/2$. From the
definitions one can derive the following relation between these two
constants:
\begin{equation}
C_{K} = \frac {3\cdot2^{1/3}\Gamma(2/3)\Gamma(d/2)}
{(d+2/3)\Gamma(d/2+1/3)}\, C_{K}' ,
\label{sviaz}
\end{equation}
which for $d=3$ gives (cf. \cite{Monin})
\begin{equation}
C_{K} = \frac{9\cdot2^{1/3}\sqrt{\pi}\Gamma(2/3)}
{22\Gamma(11/6)} \, C_{K}' = \frac{27}{55} \Gamma(1/3) C_{K}' .
\label{sviaz_1}
\end{equation}

Using the exact relation $S_{3}(r)=-12\E r/d(d+2)$ that follows from the
energy balance equation (see e.g. \cite{Monin,Legacy} for $d=3$), the
constant $C_{K}$ can be related to the inertial-range skewness factor
${\cal S}$:
\begin{equation}
{\cal S} \equiv S_{3}/S_{2}^{3/2} = -[12/d(d+2)]\,C_{K}^{-3/2}.
\label{sviaz_2}
\end{equation}

All these relations refer to real physical quantities in the inertial
range, which in the stochastic model (\ref{1.1}), (\ref{1.2})
corresponds to $\eps=2$ and $m=0$ in the random force correlator
(\ref{1.9}).

Many studies have been devoted to the derivation of $C_{K}$ within the
framework of the RG approach;
see Refs. \cite{UFN,turbo,JETP,48,50,Zhou,78,85,Giles,Jap}. In order to
obtain $C_{K}$, it is necessary to augment the solution of the RG equation
for $S_{2}$ by some formula that relates the amplitude $D_{0}$ in the random
force correlator (\ref{1.9}) to the physical parameter $\E$. In particular,
in \cite{48,50} the first-order term of the $\eps$ expansion for the
pair correlator was combined with the so-called eddy-damped quasinormal
Markovian approximation for the energy transfer function, taken at $\eps=2$.
More elementary derivation, based on the exact relation (\ref{1.3}) between
$\E$ and the
function $d_{f}(k)$ from (\ref{1.9}) was given in \cite{JETP}; see also
\cite{UFN,turbo}. In spite of the reasonable agreement with the experiment,
such derivations are not immaculate from the theoretical viewpoints. Their
common flaw is that the relation between $\E$ and $D_{0}$ is unambiguous
only in the limit $\eps\to2$ [see Eq. (\ref{2.74}) in Sec.~\ref{sec:Model}],
so that the coefficients of the corresponding $\eps$ expansions can, in fact,
be made arbitrary; see the discussion in Ref. \cite{85} and Sec.~2.10
of \cite{turbo}. The ambiguity is a consequence of the fact that the notion
itself of the Kolmogorov constant has no unique extension to the
nonphysical range $\eps<2$.

The experience on the RG theory of critical behavior shows that well-defined
$\eps$ expansions can be written for universal quantities, such as critical
exponents, normalized scaling functions and ratios of amplitudes in scaling
laws \cite{Zinn,book}. The constant $C_{K}$ extended to the range $\eps<2$
as in \cite{48,50} or \cite{JETP} involves a bare parameter, $D_{0}$, and
hence is not universal in the above sense.

To circumvent this difficulty, we propose below an alternative derivation
that relates $C_{K}$ to an universal quantity and thus does not involve any
relation between $D_{0}$ and $\E$. Consider the ratio
\begin{equation}
Q(\eps)\equiv {\cal D}_{r} S_{2}(r) / |S_{3}(r)|^{2/3}=
{\cal D}_{r} S_{2}(r) / (-S_{3}(r))^{2/3}
\label{RR}
\end{equation}
with ${\cal D}_{r} \equiv  r\partial/\partial r$. As we shall see below,
the quantity $Q(\eps)$ is universal and can be
calculated in the form of a well-defined $\eps$ expansion. On the other
hand, its value at $\eps=2$ determines the Kolmogorov constant and the
skewness factor through the exact relations
\begin{equation}
C_{K}= \left[3Q(2)/2\right]\left[12/d(d+2)\right]^{2/3},
\quad
{\cal S} = - \left[3Q(2)/2 \right]^{-3/2}
\label{trud}
\end{equation}
that follow from the definitions, relation (\ref{sviaz_2}) and the identity
${\cal D}_{r} r^{\alpha} = \alpha r^{\alpha}$ for any $\alpha$.

Solving the RG equations for the quantities in $Q(\eps)$ in the inertial
range ($\Lambda r \to \infty$ and $mr\to 0$) for general $0<\eps\le 2$ gives
(see Sec.~\ref{sec:QFT}):
\begin{equation}
S_{3}(r) = D_{0} r^{-3\Delta_{\varphi}} f_{3} (\eps), \quad
{\cal D}_{r} S_{2}(r) = D_{0}^{2/3} r^{-2\Delta_{\varphi}}
f_{2}(\eps), \quad \Delta_{\varphi} = 1-2\eps/3.
\label{RGS}
\end{equation}
It is important here that the operation ${\cal D}_{r}$ kills the constant
contribution $\langle \varphi^{2} \rangle$ in $S_{2}$ that diverges as
$\Lambda\to\infty$ for $\eps<3/2$ (see below); in $S_{3}$ such constant
contributions are absent.

It follows from Eq. (\ref{RGS}) that the ratio
$Q(\eps)=f_{2}/(-f_{3})^{2/3}$, in contrast to its numerator and
denominator, does not depend on the amplitude $D_{0}$, it is universal
(in the above sense) and can be calculated in the form of a well-defined
$\eps$ expansion. The latter has the form $Q(\eps)=\eps^{1/3} p(\eps)$,
where $p(\eps)$ is a power series in $\eps$ (see below). We shall find
the first two terms of $p(\eps)$, which corresponds to the two-loop
accuracy in representations (\ref{FPd}) and (\ref{epscal}).
In this sense, one can speak about the two-loop approximation
for the Kolmogorov constant (previous attempts have been confined
with the first order).

To avoid possible misunderstandings, we stress again that we shall not try
to extend the definition of the physical quantities $C_{K}$ and ${\cal S}$
to the whole interval $0<\eps<2$ and to construct the corresponding $\eps$
expansions from the known expansion for $Q(\eps)$. Instead, the latter is
used to give the value of $Q(2)$, which, in its turn, determines $C_{K}$
and ${\cal S}$ through the relations (\ref{trud}) that make sense only
for $\eps=2$.

Using the definition (\ref{struc}), the function $S_{2}$ can be related
to the momentum-space pair correlation function (\ref{G}) as follows:
\begin{eqnarray}
S_{2}(r) = 2\int \frac{d{\bf k}}{(2\pi)^{d}}
\,G(k)\, \left[1-({\bf k}\cdot{\bf r})^{2}/(kr)^{2}\right]
\left\{1- \exp \left[{\rm i} ({\bf k}\cdot{\bf r})\right]\right\}.
\label{atas}
\end{eqnarray}
Applying the operation ${\cal D}_{r} \equiv  r\partial/\partial r$ gives:
\begin{eqnarray}
{\cal D}_{r} S_{2}(r) = 2 \int \frac{d{\bf k}}{(2\pi)^{d}}\, G(k)\,
\left[1-({\bf k}\cdot{\bf r})^{2}/(kr)^{2}\right]\,
({\bf k}\cdot{\bf r}) \,\sin ({\bf k}\cdot{\bf r}).
\label{atas1}
\end{eqnarray}
In order to obtain the inertial-range form of the function
${\cal D}_{r} S_{2}(r)$ it is sufficient to substitute the asymptotic
expression (\ref{dimG}) into (\ref{atas1}). A straightforward
calculation gives:
\begin{eqnarray}
{\cal D}_{r} S_{2}(r) = \frac {2(d-1) \Gamma(2-2\eps/3)}
{(4\pi)^{d/2}\Gamma(d/2+2\eps/3)}\, g_{*}^{1/3}R(1,0,g_{*}) \,
D_{0}^{2/3} (r/2)^{-2\Delta_{\varphi}}
\label{atas4}
\end{eqnarray}
with the amplitude $R(1,0,g_{*})$ from (\ref{dimG}). It is important here
that the resulting integral exists for all $0<\eps<2$. It is the operation
${\cal D}_{r}$ that ensures the convergence of the integral (\ref{atas1})
with the function (\ref{dimG}); the original integral (\ref{atas}) would
be UV divergent for $0<\eps<3/2$ while the analogous expression for the pair
correlator (that is, the plain Fourier transform of (\ref{dimG})) would be
IR divergent for $\eps>3/2$.

The needed terms of the $\eps$ expansion for $f_{3}$ can be obtained
not only from the direct perturbative calculation, but also from the
exact expression
\begin{equation}
S_3(r)= -\frac{3(d-1) \Gamma(2-\eps)}
{(4\pi)^{d/2}\Gamma(d/2+\eps)} \, D_{0} (r/2)^{-3\Delta_{\varphi}}
\label{Exa}
\end{equation}
that follows from the energy balance equation and in the limit $\eps\to2$,
along with the formula (\ref{2.74}), reproduces the correct coefficient
$-12/d(d+2)$.

The pole $\sim 1/(2-\eps)$ cancels the vanishing factor $\sim (2-\eps)$
in (\ref{2.74}) and ensures the finiteness of $S_{3}$ at $\eps=2$ when
expressed in terms of the physical parameter $\E$. In the ratio (\ref{RR})
this pole must cancel with the (hypothesized) singularity in the pair
correlator; see the discussion in the end of Sec.~\ref{sec:bsk}.

Thus from Eqs. (\ref{atas4}) and (\ref{Exa}) for the ratio $Q(\eps)$ one
obtains
\begin{eqnarray}
Q(\eps)= \left[4(d-1)g_{*}\bar S_{d} /9 \right]^{1/3} A(\eps,d)
R(1,0,g_{*}),
\label{atas2}
\end{eqnarray}
where the coefficient
\begin{eqnarray}
A(\eps,d) \equiv \frac {\Gamma(2-2\eps/3)\Gamma^{1/3}(d/2)
\Gamma^{2/3}(d/2+\eps)} {\Gamma(d/2+2\eps/3)\Gamma^{2/3}(2-\eps)}=
1+O(\eps^{2})
\label{atas3}
\end{eqnarray}
has no term of order $O(\eps)$ and in our approximation can be replaced by
unity. Substituting the two-loop result (\ref{FPd}) for the coordinate of
the fixed point $g_{*}$ and the first two terms (\ref{epscal}) of the $\eps$
expansion of the amplitude $R(1,0,g_{*})$ into (\ref{atas2}) gives
\begin{equation}
Q(\eps) = (1/3) [4\eps (d+2)]^{1/3} \left[ 1+ \eps\, \left(
\lambda/3+2\alpha c_{2} \right) +O(\eps^2) \right]
\label{QQK}
\end{equation}
with $\alpha$ and $\lambda$ from Eq. (\ref{FPd}) and
$c_{2}$ from (\ref{BD}). For $d=3$ this gives
\begin{equation}
Q(\eps) = (1/3)(20\eps)^{1/3}\left[1+0.525\eps+O(\eps^2)\right].
\label{QQ}
\end{equation}

The value of $Q(\eps)$ at $\eps=2$ determines the Kolmogorov constant and
skewness factor through the exact relations (\ref{trud}), which for $d=3$
have the form $C_{K}=6\cdot 10^{-2/3}\,Q(2)$ and
${\cal S}= - [1.5 \cdot Q(2)]^{-3/2}$.
Let us denote as $C_{K}^{(n)}$ and ${\cal S}^{(n)}$ the results
obtained using the $n$-loop approximation for the RG functions and the
corresponding approximations for the scaling functions. Substituting the
value of (\ref{QQ})  with $\eps=2$ into these relations we obtain
\[C_{K}^{(2)}=3.02,\qquad {\cal S}^{(2)}= -0.15.\]
If we had retained only the first-order term in (\ref{QQ}) we would have
obtained
\[C_{K}^{(1)}=1.47, \qquad {\cal S}^{(1)}= -0.45.\]
We also recall the experimental estimates recommended in \cite{Monin}:
\[ C_{K} \approx 1.9, \qquad  {\cal S} \approx -0.28; \]
it is worth noting that
they lie in between of the first-order and second-order approximations.

Let us conclude this section with a brief discussion of the $d$-dimensional
case. Results of the two-loop calculation for several values of $d$,
including the limits $d\to\infty$ and $d\to2$, are given in
Table~\ref{table2}. It includes: the parameter $\lambda$ that determines
the second-order corrections to the coordinate of the fixed point $g_{*}$
from Eq. (\ref{FPd}) and the exponent $\omega$ from Eq. (\ref{Omega}),
the parameter $c_{2}\equiv c_{2}|_{\eps=0}$ that determines the second-order
correction to the scaling function of the equal-time pair correlator in
Eq. (\ref{epscal}), the contributions ${\cal B}$ and ${\cal D}$ from
the diagrams $R_{2,3}$ and $R_{1}$ that determine $c_{2}$ via Eq.
(\ref{BD}), and the first-order and second-order approximations for
the Kolmogorov constant $C_{K}^{(n)}$ ($n=1,2$).

One can see that, for $d\to\infty$, the two-loop contributions become
relatively smaller for all these quantities. This gives some quantitative
support to the old idea that $1/d$ can be a promising expansion parameter
for the stochastic NS problem \cite{FFR}. However, in contrast to the
Kraichnan model, where the $O(1/d)$ approximation for the anomalous
exponents is a function linear in $\eps$ \cite{Falk1}, in our case the
second-order relative corrections have a finite limit at $d\to\infty$,
so that the $O(1/d)$ approximation for $g_{*}$, $\omega$, $Q(\eps)$
are infinite series in $\eps$. On the other hand, like for the Kraichnan
model, the calculation of the diagrams drastically simplifies for
$d\to\infty$: the leading terms can be calculated analytically. We
believe that the three-loop calculation is a feasible task at $d=\infty$;
this work is now in progress.

For $d\to\infty$, the constant $C_{K}$ decreases as $C_{K} \propto 1/d$;
see Table~\ref{table2}.
From Eq. (\ref{sviaz}) one then obtains $C_{K}' \propto d^{1/3}$ for the
coefficient in the energy spectrum, in agreement with the earlier results
obtained within the DIA \cite{FFR} or the RG \cite{UFN,turbo}.

For $d\to2$, the two-loop contributions diverge; this is a manifestation
of the additional UV divergence that emerge in the 1-irreducible function
$\langle\varphi'\varphi'\rangle$ at $d=2$. In this region our results
cannot be trusted, and additional renormalization should be performed in
order to remove the divergences at $d=2$; see e.g. Sec.~3.10 of Ref.
\cite{turbo} and references therein. This is necessary, in particular, in
the discussion of possible crossovers in the inertial-range behavior that
can occur between $d=3$ and 2; see Refs.~\cite{FF}.
It is worth noting that, for all $d$, the main contribution to the
coefficient $c_{2}={\cal D}+2{\cal B}$ in (\ref{BD}) comes from the
term ${\cal D}$, that is, from the diagram $R_{1}$. What is more, the
contribution from ${\cal B}$ totally vanishes at $d=\infty$ while the
contribution from ${\cal D}$ diverges at $d=2$. It thus might happen
that the aforementioned additional renormalization is needed to obtain
more accurate numerical predictions in three dimensions; we shall
consider this important problem elsewhere.

\section{Conclusion} \label{sec:Con}

We have accomplished the complete two-loop calculation of the renormalization
constant and RG functions for the stochastic problem (\ref{1.1})--(\ref{1.9})
and derived the coordinate of the fixed point, the UV correction exponent
$\omega$, the Kolmogorov constant $C_{K}$ and the inertial-range skewness
factor ${\cal S}$ to the second order of the corresponding $\eps$ expansions.
The new point is not
only the inclusion of the second-order correction, but also the derivation
of $C_{K}$ through the universal (in the sense of the theory of critical
behavior) quantity (\ref{RR}).

Of course, one should have not expected that the second-order terms of the
$\eps$ expansions would be small in comparison to the first-order terms.
The experience on the RG theory of critical behavior shows that such
corrections are not small for dynamical models (in contrast to static ones)
and for amplitudes (in contrast to exponents); see \cite{Zinn,book}.
It is thus rather surprising that in our case the account of the two-loop
contributions leads to reasonable changes in the results.

Although the $\eps^2$ correction to $\omega$ in (\ref{Omega}) is rather
large, it does not change its sign and hence does not destroy the IR
stability of the fixed point for the real $\eps=2$.

The first-order approximation $C_{K}^{(1)}=1.47$ underestimates, and
the second-order approximation $C_{K}^{(2)}=3.02$ overestimates the
conventional experimental value of the Kolmogorov constant
$C_{K} \approx 1.9$ \cite{Monin}.
Thus the experimental value of $C_{K}$ (and hence for ${\cal S}$) lies
in between of the two consecutive approximations. A similar situation is
encountered for the well-known Heisenberg model \cite{Monin}, where the
analog of the Kolmogorov constant is known exactly and lies between
the first-order and second-order approximations given by the corresponding
$\eps$ expansion \cite{Hei}. If we assume, by the analogy with the
Heisenberg model, that the (unknown) exact predictions for $C_{K}$ and
${\cal S}$ lie between the first two approximations, we may conclude
that our calculation has given satisfactory estimates
for these quantities.

The two-loop corrections become relatively small for $d\to\infty$, which
confirms the relevance of the $1/d$ expansion for the issue of
inertial-range scaling behavior.

One may thus conclude that our results confirm the applicability of the
RG formalism and the $\eps$ expansion to the calculation of important
characteristics of the real fluid turbulence. The next important steps
should be the derivation of the two-loop results for the case of
a passive scalar advected by stochastic NS equation; inclusion of the
additional renormalization near two dimensions, and the three-loop
calculation in the limit $d\to\infty$; this work is now in progress.

\acknowledgments
The authors thank Michal Hnatich and Juha Honkonen for discussions.
The work was supported by the Nordic Grant for Network Cooperation
with the Baltic Countries and Northwest Russia No.~FIN-18/2001, the
GRACENAS Grant No.~E00-3-24, and the program ``Universities of Russia.''
N.V.A. was also supported by the Academy of Finland (Grant No.~79781).

\appendix
\section*{}

Below we give explicit expressions for the integrands $f_{0}$--$f_{4}$
in the expressions (\ref{A17}) for the two-loop diagrams of the
1-irreducible correlation function $\langle\varphi'\varphi\rangle$.
They have the form
\[ f_i(z,x) = x^{2} (1-z^2)^{(d-1)/2}\,
\frac {S_{d-1}} {64(d-1)d(d+2)S_{d}}\, \bigl[
\psi_{i}(z,x)+\psi_{i}(-z,x) \bigr] \]
with $S_{d}$ from (\ref{surface}) and $x\equiv k_{1}/k_{2}$.
Note that the functions $f_i(z,x)$ are even in $z$ [see the remark
below Eq.~(\ref{A17})]. The functions $\psi_{i}\equiv\psi_{i}(z,x)$ in $d$
dimensions have the forms (the function $\psi_0$ is even in $z$,
while the other functions $\psi_i$ are not)
\begin{eqnarray}
\psi_0 &=& \big\{({x}^{20}+1) \big[ ( - 3{d}^{3} + 8 {d}^{2} -
7{d} - 16 ) + ( - 14 {d}^{2} + 100 {d} + 40 ) {z}^{2} - 120 {d}
{z}^{4} \big]
\nonumber \\ &+& ({x}^{18}+x^2) \big[ ( - 30
{d}^{3} + 76 {d}^{2} - 78 {d} - 64 )  + ( 12 {d}^{3} + 24 {d}^{2}
+ 552 {d} - 144 )
 {z}^{2}
\nonumber \\ &+&   ( -184 {d}^{2} - 76 {d} + 184 ) {z}^{4}
 -440 {d} {z}^{6} \big]
\nonumber \\ &+& ({x}^{16}+x^4) \big[ ( - 135 {d}^{3} +
328 {d}^{2} - 395 {d} + 48 ) + ( 96 {d}^{3 } +522 {d}^{2} + 1080
{d} - 1272  ) {z}^{2} \nonumber \\ &+&
 ( 63 {d}^{3} -1796 {d}^{2} +1495 {d} + 1728
 ) {z}^{4} + (  1054
{d}^{2} -2984 {d} - 576 ) {z}^{6} + 816 {d} {z}^{8} \big]
\nonumber
\\ &+& ({x}^{14}+x^6) \big[ ( - 360 {d}^{3} + 848 {d}^{2} -  1160{d} +
768 ) + ( 336 {d}^{3} +1952 {d}^{2} - 88    {d} - 2752
 ) {z}^{2}
 \nonumber \\ &+& ( 378 {d}^{3} -7348 {d}^{2} + 5918  {d} + 3136
 ) {z}^{4} + ( - 312 {d}^{3
} + 6124 {d}^{2} - 5320  {d} - 1232
 ) {z}^{6}
 \nonumber \\ &+& ( -1432 {d}^{2} -  208 {d} + 224 ) {z}^{8}
 + 1248 {d} {z}^{10} \big]
 \nonumber \\ &+& ({x}^{12}+x^6)   \big[( - 630 {d}^{3}
+ 1456 {d}^{2} - 2158 { d} + 2016)
              + ( 672 {d}^{3} + 3844{d}^{2} - 3740 {d} - 2864
 ) {z}^{2}
 \nonumber \\ &+& ( 945 {d} ^{3} - 16172{d}^{2} + 10945 {d} -
64) {z}^{4} + ( - 1248{d}^{3 } +15362 {d}^{2} -  2296 {d} + 1408 )
{z}^{6}
\nonumber \\ &+&  (240 {d} ^{3} - 5376 {d}^{2} -  4224 {d}
-  768 ) {z}^{8} + ( 832 {d}^{2}
 + 1792 {d} + 128 ) {z}^{10} - 640 {d} {z}^{12}\big]
 \nonumber\\&+&
  2{x}^{10} \big[ ( - 378 {d}^{3} + 868 {d}^{2} - 1322 {d} + 1344
 )  + ( 420 {d}^{3} + 2376 {d}^{2} - 3024 {d} - 1200 )
  {z}^{2} \nonumber\\&+&
      ( 630 {d}^{3} - 10436 {d}^{2}
+ 6478  {d} - 1656 ) {z}^{4}
         + (  - 936 {d}^{3} + 10292 {d}^{2} + 416 {d} + 2064
  ) {z}^{6} \nonumber\\ &+&  ( 240 {d}^{
3} - 3912 {d}^{2} - 3360 {d} - 480) {z}^{8}  + ( 832 {d}^{2} + 544
{d} ) {z}^{10} + (  - 128 {d}^{2} + 256 {d} ) {z}^{12} \big]
\big\} \nonumber\\ &\big/& \left[
 ({x}^{2} + 2 {x} {z} + 1 )^{3}
  ({x}^{2} +  {x} {z} + 1 )^{3} ({x}^{2} - 2 {x} {z} + 1 )^{3}
   ({x}^{2} - {x}
 {z} + 1 )^{3}
  \right];
\nonumber\\  {{\psi}_{1}} &=& 2 \big \{ {x}^{7} {z} (  - 2 {d}^{2} +
2 {d} ) + {x}^{6} \big[ ( {d}^{3} - 4 {d}^{2} + {d} + 6 ) + ( -
6{d}^{2} + 6{d} - 12){z}^{2} \big] + {x}^{5}{z} \big[ ( 5{d}^{3} -
20{d}^{2} + 7{d} - 4 ) \nonumber\\ &+&  ( - 4{d}^{2} + 4 {d}
 - 16 ) {z}^{2}  \big] + {x}^{4} \big[ ( 3 {d}^{3} - 10 {d}^{2} + 3 {d} )
  + ( 8 {d}^{3} - 26 {d}^{2} + 10{d} - 28 )
{z}^{2}\big] \nonumber\\ &+& {x}^{3}{z} \big[ (10 {d}^{3} - 28
{d}^{2} + 10 {d} - 20 )
   + ( 4 {d}^{3} - 8 {d}^{2} + 4 {d} ) {z}^{2} \big] + {x}
   ^{2}\big[
 ( 3 {d}^{3} - 8 {d}^{2} + 3 {d} - 6 )  \nonumber\\ &+&
 ( 8 {d}^{3} - 16 {d}^{2} + 8 {d} )
 {z}^{2}\big]
      + {x}{z} ( 5 {d}^{3} - 10 {d}^{2} + 5 {d} )
     + {d} ^{3} - 2 {d}^{2} + {d}
 \big \} \nonumber\\ &\big/&
   \left[
   ({x}^{2}+ 2{x}{z}  + 1 )^{2}
 ({x}^{2} + {x}{z}+ 1 )^{2} \right];
\nonumber\\  {{\psi}_{2}} &=& 2\big\{ {x}^{4}( 2{d}^{2}
 - 2{d} + 2) + {x}^{3}{z}( - 4{d}^{2} + 4{d
} - 4) + {x}^{2} \big[(5{d}^{2} - 5{d} + 6) + ( 2{d}^{2} - 2{d} +
2){z}^{2}\big] \nonumber\\ &+& {x}{z} ( - 5{d}^{2} + 5{d} - 6)
 + 3{d}^{2} - 3{d} + 6  \big\} \big[ 2{x}^{3}{z} + {x}^{2}( {d} - 3) +
{x}{z}(- 2{d} + 2)
  + {d} - 1 \big]
\nonumber\\ &\big/&   \left[({x}^{2} -2{x}{z} + 1 ) ({x}^{2} -
{x}{z} + 1)^{3}
 \right];
\nonumber\\ {{\psi}_{3}} &=&  2 d(d-1)\, \big[ 2{x}^{3}{z} + {x}^{2}
( {d} - 3 ) + {x} {z}(-2{d} + 2)  + {d} - 1 \big] \big/
\big[({x}^{2} -2{x}{z} + 1 ) ({x}^{2} -{x}{z} + 1 ) \big];
\nonumber\\ {{\psi}_{4}} &=& \big\{({x}^{4}+1)(
 3 {d}^{2} - 3 {d} + 2 )  + ({x}^{3}+x){z} ( - 7 {d}
^{2} + 7 {d} - 8 )  + 2 {x}^{2} \big[ ( 3 {d}^{2} - 3 {d} + 2 )
 + ( 2 {d}^{2} - 2 {d} + 4 ) {z}^{2}\big]\big\}
\nonumber\\
  &\big[& ({x}^{4}+1)( - {d} + 1 )  +({x}^{3}+x){z}( 2
{d} - 4 )  + 2{x}^{2}( - {d} + 3 )  \big]
 \big/ \left[
  ({x}^{2} -2 {x} {z}
+ 1 )^{3} ({x}^{2} -{x} {z} + 1)^{3}
 \right].
\nonumber
\end{eqnarray}

\begin{table}
\caption{Results of the two-loop calculation for different values of
$d$.}
\label{table2}
\begin{tabular}{cccccc}
{} & $d=2+2\delta$ & $d=2.5$ & $d=3$  & $d=5$ &  $d\to\infty$ \\
\tableline
$\lambda$   & $-1/3\delta+O(\delta^{0})$  & $-2.296$  & $-1.101$
& $-0.560$ & $-1/3+O(1/d)$  \\
${\cal B}$  & $O(\delta^{0})$ & 0.0013 & $-0.000057$  & -0.00194
&  $O(1/d)$\\
${\cal D}$  & $1/64\delta+O(\delta^{0})$ & 0.0999 & $0.06699$  &
0.0436 &  $1/32+O(1/d)$\\
$c_{2}$     & $1/64\delta+O(\delta^{0})$  & 0.103  & $0.0669$  &
0.0397 &  $1/32+O(1/d)$ \\
$C_{K}^{(1)}$ & $2.08+O(\delta)$ & 1.72 & 1.47  & 1.35 &
$5.24/d+O(1/d^2)$ \\
$C_{K}^{(2)}$ & $0.93/\delta+O(\delta^{0})$ & 4.74 & 3.02  & 1.84 &
$5.82/d+O(1/d^2)$ \\
\end{tabular}
\end{table}

\end{document}